\documentclass[a4paper,fleqn,usenatbib]{mnras}

\usepackage{newtxtext,newtxmath}

\usepackage[T1]{fontenc}
\usepackage{ae,aecompl}

\usepackage{graphicx}	
\usepackage{amsmath}	
\usepackage{amssymb}	

\title[Orbital period modulation in close binaries]{Internal magnetic fields, spin-orbit coupling, and orbital period modulation in close binary systems}

\author[A. F. Lanza]{
A. F. Lanza,\thanks{E-mail: antonino.lanza@inaf.it}
\\
INAF-Osservatorio Astrofisico di Catania, Via S.~Sofia,78 - 95123 Catania, Italy\\
}

\date{Accepted XXX. Received YYY; in original form ZZZ}

\pubyear{2019}

\begin{document}
\label{firstpage}
\pagerange{\pageref{firstpage}--\pageref{lastpage}}
\maketitle

\begin{abstract}
We introduce a new model to explain the modulation of the orbital period observed in close stellar binary systems based on an angular momentum exchange between the spin of the active component and the orbital motion. This spin-orbit coupling is not due to tides, but is produced by a non-axisymmetric component of the gravitational quadrupole moment of the active star due to a persistent non-axisymmetric  internal magnetic field. The proposed mechanism easily satisfies all the energy constraints having an energy budget  $\sim 10^{2}-10^{3}$ times smaller than those of previously proposed models and is supported by the observations of persistent active longitudes in the active components of close binary systems. We present preliminary applications to three well-studied  binary systems to illustrate the model. The case of stars with hot Jupiters is also discussed  showing that no significant orbital period modulation is generally expected on the basis of the proposed model. 
\end{abstract}




\section{Introduction}
The orbital period $P$ of an eclipsing binary can be measured with  high precision thanks to the periodic character of the orbital motion. Long-term monitoring  led to the discovery of cyclic modulations of the orbital periods of close binaries with at least one late-type component star, that is, with a spectral type later than $\sim$ F5 \citep{Hall89,Hall90}. In Algol and RS Canum Venaticorum systems, the typical relative amplitudes are $\Delta P/P \sim (1-3) \times 10^{-5}$ with a typical  modulation period $P_{\rm mod} \sim 30-50$ years, while in more compact systems, such as Cataclysmic Variables (CVs), post-common envelope binaries (PCEBs), and contact binaries of the W Ursae Maioris class, typical $\Delta P/P \sim  (0.1 - 3) \times 10^{-6}$ with $P_{\rm mod}$  ranging from several years to a few decades \citep[e.g.,][and references therein]{LanzaRodono99}. Low-mass X-ray binaries and millisecond binary pulsars also share a similar phenomenology \citep[e.g.,][]{Wolffetal09,Lazaridisetal11,PletschClark15}. 

In these binary systems, the late-type secondaries have an outer convective zone and are rotating fast owing to the strong tidal interactions with their companions. Therefore, the  basic ingredients for a vigorous hydromagnetic dynamo action are in place, leading Hall to conjecture that the orbital period modulation is somehow associated with the hydromagnetic dynamo action in the secondary components of close binaries. 

Several models have been proposed to account for this connection, in particular those based on a cyclic variation of the gravitational quadrupole moment of the secondary components, originally proposed by  \citet{MateseWhitmire83} and linked to the dynamo action by \citet{ApplegatePatterson87}, \citet{Applegate92}, and \citet{Lanzaetal98}. By modulating the gravitational quadrupole moment of the active star, the orbital motion of the companion is instantaneously perturbed without requiring any exchange of angular momentum between stellar spin and the orbit. Specifically, when the quadrupole moment increases, the gravitational field in the equatorial plane of the secondary increases, thus forcing the companion to move closer and faster than during the phases when the quadrupole  decreases. Changing the quadrupole moment requires a change in the internal density distribution of the secondary star, that implies a direct perturbation of the internal hydrostatic balance by the magnetic fields as in \citet{ApplegatePatterson87}, or an indirect effect produced by redistributing the internal angular momentum which changes the centrifugal force as in \citet{Applegate92}.  By including both the effects of the Lorentz and centrifugal forces, \citet{Lanzaetal98} and \citet{LanzaRodono99} showed that the energy required to produce a given change of the quadrupole moment can be reduced by a factor of $\sim 2$ with respect to the original Applegate's model. 

These models have been criticized because the modulation of the  quadrupole moment requires more energy than is available from the stellar luminosity during the duration of the cycle. \citet{MarshPringle90} reached this conclusion  for the mechanisms invoking a direct perturbation of the hydrostatic balance by the Lorentz force, while \citet{Lanza05,Lanza06} showed that the amplitude of the required differential rotation changes in the \citet{Applegate92} model is significantly larger than the variations observed in RS CVn systems and the energy dissipated by the shear during the cycles exceeds that available from the stellar luminosity by at least two orders of magnitude. More recent studies, based on refinements of the  Applegate's or Lanza's approaches have confirmed these results showing that these mechanisms can be viable, in the best case, only for a restricted range of parameters of the close binary systems \citep[e.g.,][]{Brinkworthetal06,Volschowetal16,Navarreteetal18,Volschowetal18}. 

The investigation of eclipse time changes in PCEBs has recently become of relevant interest because interpreting the phenomenon as a light-time effect  leads to masses of the third body in the giant planet or brown dwarf regimes. Subsequent investigations of the dynamical stability of the systems showed that in general the orbits of those third bodies are unstable, an exception being the candidates proposed  around NN Serpentis \citep{Boursetal16}. The possibility that the Applegate mechanism can induce  variations in the times of mid-transits in systems with close-by planets has also been proposed \citep{WatsonMarsh10}, thus making models to explain orbital period modulation worth of further investigation. 

In this work, an alternative mechanism to explain orbital period modulation in close binaries with late-type components is proposed based on a permanent non-axisymmetric  gravitational quadrupole moment.  Such a quadrupole moment  is produced by non-axisymmetric internal magnetic fields in the convection zone of the active component. The model energetic requirements  are shown to be fully compatible with the  stellar luminosity and the observed timescales. 

\section{Model}
\label{model_desc}
\subsection{Overview}
\label{model_overview_text}
\begin{figure}
\centerline{
\includegraphics[height=10cm,width=7cm,angle=270]{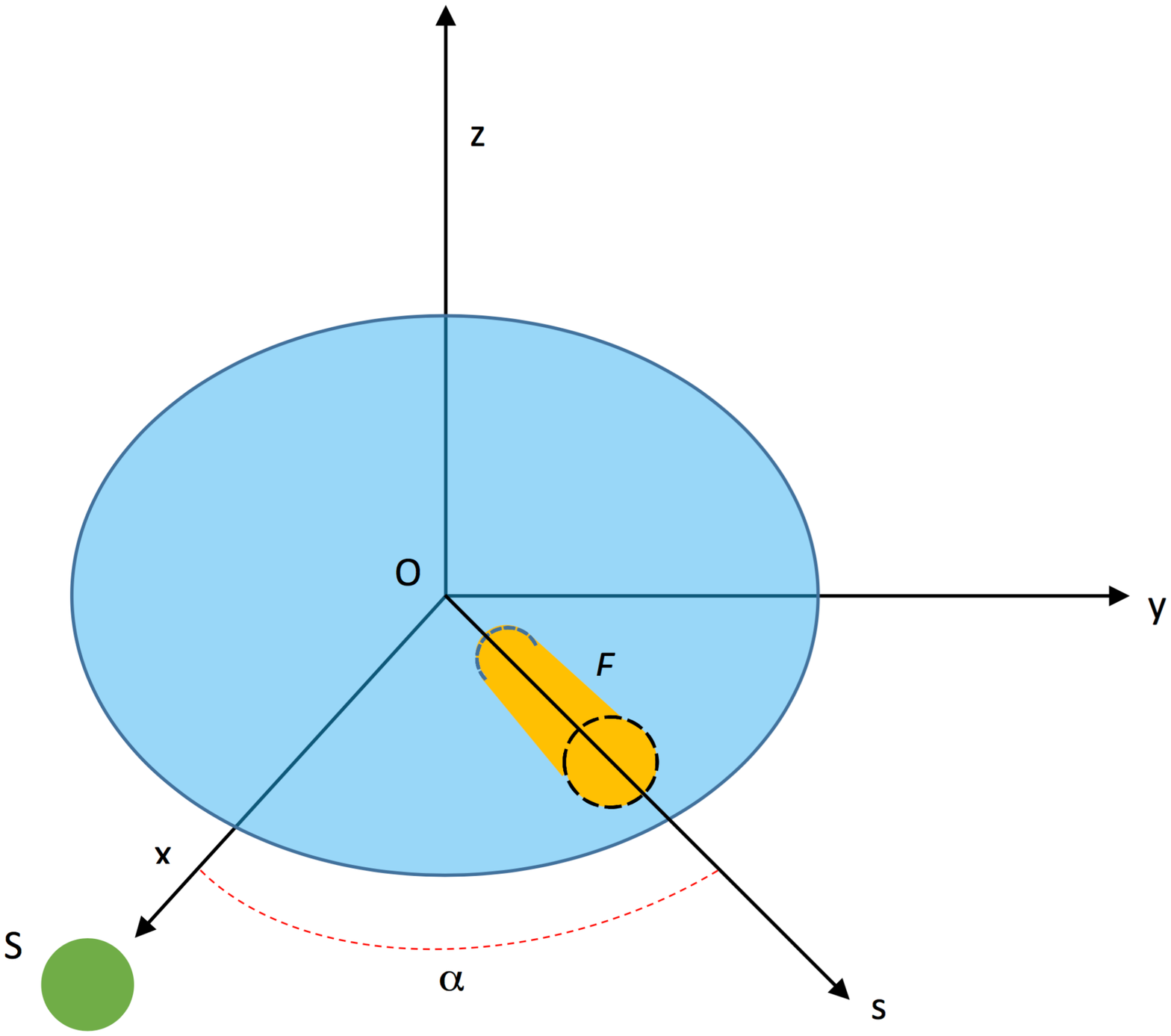}} 
\caption{Sketch of a close binary system with a magnetic flux tube $F$ inside the active component star (depicted in orange). The secondary star  $S$ is considered as a point mass and is rendered in green. The angle $\alpha$ between the line joining the centres of the two stars and the axis of the flux tube, assumed to lie in the equatorial plane,  is indicated. }
\label{model_overview}
\end{figure}

In this subsection, we provide a qualitative description of our model deferring quantitative considerations to the next subsections. In Fig.~\ref{model_overview}, we consider a Cartesian reference frame with the origin $O$ in the barycentre of the magnetically active star and the $\hat{z}$ axis along its spin axis, while the $\hat{x}$ axis is directed along the line joining the centres of the two components in the equatorial plane. We consider a radial magnetic flux tube $F$ (in orange) in the equatorial plane inside the convection zone of the active star. The magnetic pressure contributes to the pressure balance inside the flux tube thus reducing the density of the plasma inside it. Therefore, the outer gravitational field of the active star is modified by the presence of the flux tube because of this density perturbation. The orbital motion of the companion $S$, considered as a point mass orbiting in the equatorial plane, is affected and the effect depends on the angle $\alpha$ between the axis $\hat{x}$ joining the centres of the two stars and the axis $\hat{s}$ of the flux tube. 

In the case of perfectly rigid rotation and tidal synchronization of the two components, $\alpha$ would stay constant, but, if the system is not perfectly synchronized, $\alpha$ will vary in time producing a time-dependent effect on the orbit of the companion. The period of the modulation of the orbital period will be the period of the variation in the angle $\alpha$, while  the amplitude will depend on the strength of the magnetic field inside the flux tube. 

\begin{figure}
\centerline{
\includegraphics[height=8cm,width=9cm,angle=0]{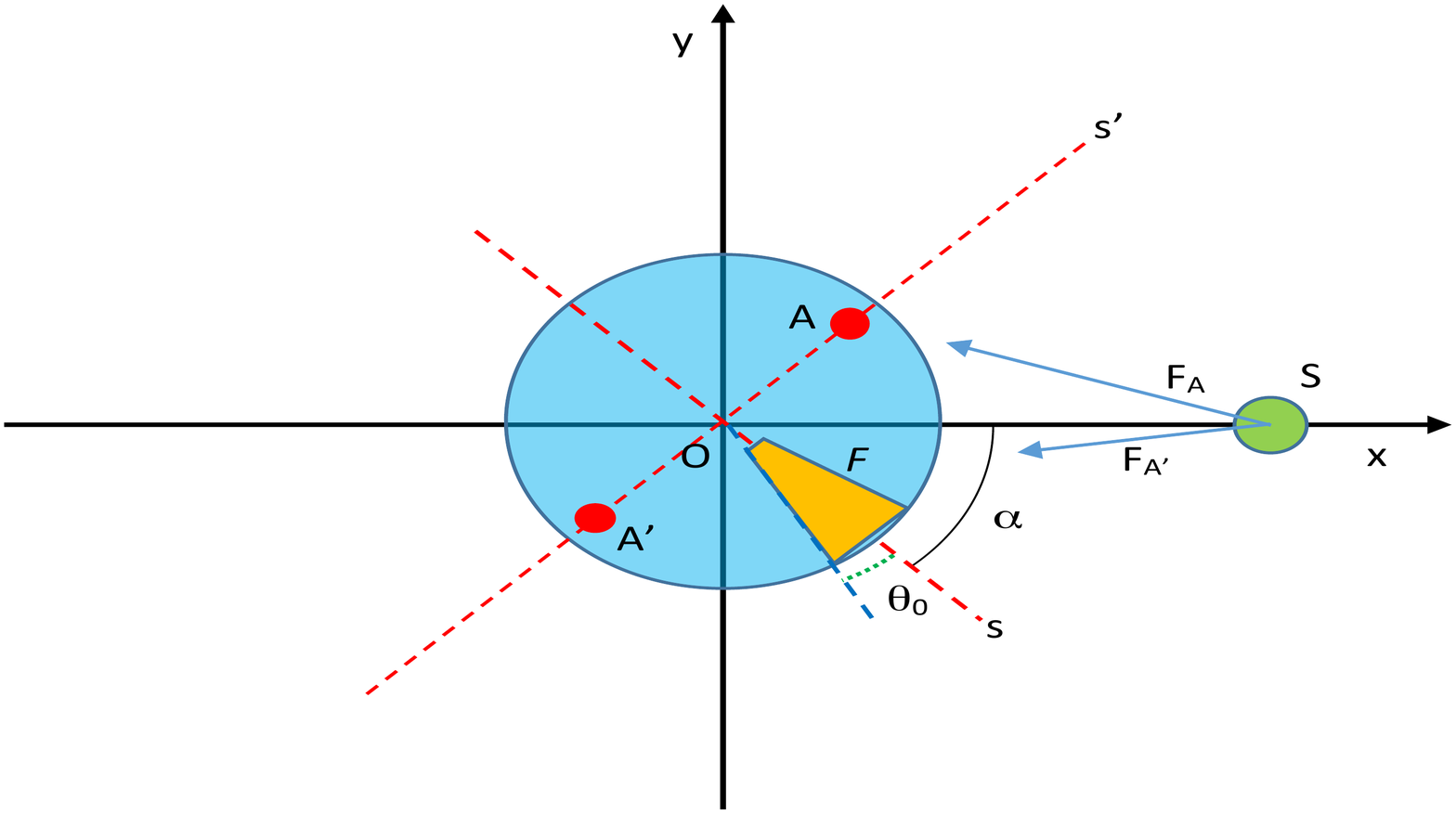}} 
\caption{Illustration of an active star with a non-axisymmetric quadrupole moment and the gravitational forces acting on the companion $S$ in a binary system.  The section of the radial flux tube having an angular radius $\theta_{0}$  is rendered in orange (see text for explanation). }
\label{quadrupole_overview}
\end{figure}

In this scenario, the non-axisymmetric component of the quadrupole moment of the active star, associated with the density perturbation inside the flux tube,  produces a torque on the orbit, thus exchanging angular momentum between the orbit and the spin of the active component. A simple representation of a star with a non-axisymmetric quadrupole moment is sketched in Fig.~\ref{quadrupole_overview} where two point masses $A$ and $A^{\prime}$ are added in the equatorial plane $xy$  of an otherwise spherically symmetric mass distribution \citep[cf.][]{MurrayDermott99}. The principal axes of inertia of this configuration are the line joining the two point masses $A A^{\prime} \equiv \hat{s}^{\prime}$, the line $\hat{s}$ orthogonal to $\hat{s}^{\prime}$ in the equatorial plane, and the axis $\hat{z}$ that is orthogonal to the equatorial plane and directed along the line of sight. The $\hat{s}$ axis is directed along the axis of the vertical flux tube $F$; its internal density is lower and the removed mass has been redistributed in the two point masses $A$ and $A^{\prime}$ displaced along  the direction $\hat{s}^{\prime}$ perpendicular to $\hat{s}$. The moment of inertia $I^{\prime}$ about the $\hat{s}^{\prime}$ axis  is minimum because the point masses lie  along the axis, while the moment $I$ about the axis $\hat{s}$ is maximum because the distance of the point masses from the axis is maximal. The non-axisymmetric quadrupole moment of this configuration is given by $T = I - I^{\prime} $ (cf. Sect.~\ref{eqs_of_motion}). The gravitational forces exerted by the two point masses $A$ and $A^{\prime}$ on the companion $S$ produce a net torque that accelerates its orbital motion exchanging angular momentum with the spin of the active component. This angular momentum exchange is periodic because it depends on the angle $\alpha$ that varies  periodically (see Sect.~\ref{eqs_of_motion}) and is responsible for the orbital period modulation of the binary system. 

We stress that this spin-orbit coupling is not produced by the tidal bulge, the deviation of which from the line joining the centres of the two stars is very small (cf. Sect.~\ref{tides_winds}), but by the non-axisymmetric component of the density perturbation that can make a large angle $\alpha$ with the line joining the centres of the two components. This allows a much faster exchange of angular momentum in spite of the modest amplitude of the density perturbation (cf. Sects.~\ref{order_of_magnitude} and~\ref{eqs_of_motion}). 

As we shall see in Sect.~\ref{applications}, a field strength of the order of $0.5-10$~T is required to account for the observed amplitude of the orbital period modulation assuming that the internal magnetic field consists of a single flux tube with a cross section area of the order of 10 percent of the total area at the base of the convection zone. Such field strengths have been obtained in magnetohydrodynamic numerical models of the dynamo in active stars, even without an overshoot layer at the base of the convection zone where strong fields can be stored \citep[][]{Browning08,Browningetal16,BrunBrowning17}. In the Sun, fields up to 10~T in the overshoot region have been considered to account for the properties of sunspot groups \citep[][]{Caligarietal95,Moreno-Insertisetal95}. Such strong fields are highly buoyant in the superadiabatic convection zone, thus we assume that they are organized in vertical (radial) magnetic flux tubes going from the base of the convection zone up to the surface where they appear as starspots. Admittedly, the presence of such large vertical magnetic flux tubes is not generally seen in present dynamo models. In their numerical simulations, \citet{Nelsonetal13} found mainly toroidal and axisymmetric fields in the bulk of the convection zone that became increasingly amplified developing intermittency and non-axisymmetric loops as the Taylor number (a measure of the influence of rotation) was increased. Those loops could then emerge producing flux tubes with a remarkable non-axisymmetric distribution \citep{Nelsonetal14}, but capturing  the full process is still beyond the possibilities of present simulations.  A tendency for non-axisymmetric fields to become  dominant with rotation rates exceeding a few times the solar angular velocity has been found also in the simulations by \cite{Vivianietal18}. Our assumption of a single vertical magnetic flux tube is adopted to simplify the computation of the density perturbation inside the magnetic structure. As a matter of fact, what is really needed is a strongly non-axisymmetric field configuration in the convection zone of the active components as suggested by such models. Nevertheless, even the most advanced simulations are still several orders of magnitude far from the magnetohydrodynamic regimes characteristic of real active stars, therefore they results should always be taken with great caution. 

In very active stars, the non-axisymmetric distribution of the photospheric magnetic fields is revealed by the persistent active longitudes for the appearance and evolution of starspots \citep[cf.][]{Lehtinenetal16}. In close binary systems, such as the prototype RS~CVn or HR~1099, a main active longitude is generally present and persists for several decades, that is, for timescales comparable with the total extension of the available observations \citep[][]{Rodonoetal95,Lanzaetal06}. Therefore, the observations  are in favour of our hypothesis that non-axisymmetric internal magnetic fields  are present in the active components of close binary systems and remain stationary for timescales  longer than the orbital period modulation cycle. Note that individual spots can form and decay on timescales much shorter than the modulation cycle, but the active longitude is a persistent feature with a long lifetime, thus we can assume that the non-axisymmetric field configuration is stationary over very long timescales. However, a word of caution is in order here because the presence of non-axisymmetric fields in the bulk of the convection zones of active components, required to produce a sufficient density perturbation in their interiors, cannot be demonstrated by these observations. The active longitudes where spots preferentially appear could be a surface phenomenon related to the concentration of photospheric fields by large-scale non-axisymmetric convective flows that have been observed in hydrodynamic simulations of rapidly rotating convection zones \citep{Brownetal08,Brunetal17}. It is interesting to note that these large-scale convective flows are present in spite of the increasing radial shear $\Delta \Omega$ in the stellar angular velocity with increasing rotation rate $\Omega$ \citep{Brunetal17}, thus withstanding the effects of differential rotation that tends to erase non-axisymmetric structures. The same is true for non-axisymmetric magnetic fields in the case of models with relative differential rotation amplitudes  $\Delta \Omega / \Omega \sim 0.1$ \citep[e.g.,][]{Vivianietal18}, although our simplifying assumption of a single radial magnetic flux tube, strictly speaking, is untenable in the case of a large radial shear.


\subsection{Order-of-magnitude estimates}
\label{order_of_magnitude}
A simple order-of-magnitude estimate of the amplitude of the orbital period modulation produced by a given non-axisymmetric quadrupole moment $T$ can be obtained by computing the torque associated with the sum of the two forces $F_{\rm A}$ and $F_{\rm A^{\prime}}$ in Fig.~\ref{quadrupole_overview} as shown by, for example, \citet{MurrayDermott99} (see their Sect.~5.3). Here we make use of the equation of motion for the true anomaly $f$ (see the third of equations~\ref{eqs_motion} in Sect.~\ref{eqs_of_motion}) that we rewrite in order of magnitude for a circular orbit ($\dot{r} = 0$) as
\begin{equation}
mr^{2} \ddot{f} \approx \frac{3Gm_{\rm S} T}{4 r^{3}},
\label{mean_motion_var}
\end{equation}
where $G$ is the gravitation constant, $m=Mm_{\rm S}/(M+m_{\rm S})$ the reduced mass of the binary, $M$ the mass of the active star, $m_{\rm S}$ the mass of the companion star,  $r$ the radius of the orbit; and we have approximated $\sin 2 \alpha \approx 1/2$. If the cycle of the orbital period modulation has a duration $P_{\rm mod}$, assuming a nearly sinusoidal modulation of the true anomaly  with respect to an unperturbed orbit, we have:
\begin{equation}
\ddot{f} = \frac{2\pi}{P_{\rm mod}} \Delta \dot{f}, 
\label{ddotf}
\end{equation}
where $\Delta \dot{f}$ is the variation of the orbital mean motion. To evaluate $\Delta \dot{f}$  for a circular orbit, we note that $ \dot{f} = n = 2\pi/P$, where $P$ is the orbital period and $n$ the mean orbital motion. By differentiating this expression, we find $\Delta \dot{f} = -n (\Delta P / P)$. Making use of equation~(\ref{ddotf}) and the Kepler III law, we recast equation~(\ref{mean_motion_var}) as:
\begin{equation}
\frac{T}{I_{\rm p}} \approx \frac{4}{3} \left(\frac{M_{\rm T}}{m_{\rm S}} \right) \left( \frac{m r^{2}}{I_{\rm p}}\right) \left( \frac{P}{P_{\rm mod}} \right) \left| \frac{\Delta P}{P} \right|,
\end{equation}
where $I_{\rm p}$ is the moment of inertia of the active star about its spin axis and $M_{\rm T}=M + m_{\rm S}$ the total mass of the binary system. For a typical RS CVn system, $mr^{2}/I_{\rm p}$ ranges between 25 and 100; assuming $M_{\rm T}/m_{\rm S} = 2$, $P = 3$~days, $P_{\rm mod} = 40$ yrs, and $\Delta P /P = 10^{-5}$, we obtain $T/I_{\rm p} = (0.7-2.7) \times 10^{-7}$. For comparison, the variation of the axisymmetric  quadrupole moment $\Delta Q$ considered by the models of \citet{Applegate92} or \citet{Lanzaetal98} is 
\begin{equation}
\frac{\Delta Q}{I_{\rm p}} = \frac{1}{9}  \left(\frac{M_{\rm T}}{m_{\rm S}} \right) \left( \frac{mr^{2}}{I_{\rm p}}  \right) \left| \frac{\Delta P}{P} \right|,
\end{equation}
that is larger by a factor $\sim P_{\rm mod}/(12P) \sim 100-1000$ than the non-axisymmetric quadrupole moment assumed in the present model. Since the quadrupole moment perturbations are directly proportional to the magnetic energy \citep{LanzaRodono99}, the present model has a strong advantage over previous models from an energetic point of view. 

\subsection{Internal magnetic fields in the active components of close binary systems}
\label{strong_fields}

We consider a spherical polar coordinate system in the reference frame introduced in Fig.~\ref{model_overview}. The distance from the barycentre $O$ of the magnetically active star is the radial coordinate $r$, the colatitude $\theta$ is measured from the North pole, and the azimuthal coordinate around the $\hat{z}$ axis is indicated with $\phi$. 
For the sake of simplicity, we describe the stationary non-axisymmetric field configuration as a single magnetic flux tube. 
The condition that the field be mainly vertical in this flux tube can be expressed as $ B_{r} \gg B_{\theta}, B_{\phi}$. Since the magnetic field is solenoidal $\nabla \cdot {\bf B} = 0$ and  we have $B_{r} r^{2} = const.$ that can be used to compute the magnetic field strength as a function of the radial coordinate $r$. Considering a magnetic flux tube extending from the base of the convection zone $r_{\rm b}$ up to the photosphere at the star radius $R$, this implies:
 \begin{equation}
 B(r, \sigma) = B_{0} (\sigma) \left( \frac{r}{r_{\rm b}} \right)^{-2},
 \label{field_strength}
 \end{equation}
 where $\sigma$ is the distance from the axis $\hat{s}$ of the vertical flux tube on a surface of constant radius $r$ and $B_{0}$ the field at the base of the stellar convection zone, that is, $B_{0}(\sigma) = B(r_{\rm b},\sigma)$. Given that the field is vertical, we can neglect the magnetic tension force  and write the pressure balance across the section of the flux tube as:
\begin{equation}
 p_{\rm i}(r,\sigma) + \frac{B^{2}(r,\sigma)}{2\mu}  = p_{\rm e}(r), 
\label{pressure_balance}
\end{equation}
where $p_{\rm i}$ is the pressure of the plasma inside the flux tube and $p_{\rm e}$ the unperturbed pressure outside the tube that, for the sake of simplicity, we assume to depend only on the radial coordinate $r$. Equation (\ref{pressure_balance}) is valid only at a sufficiently large depth where $p_{\rm e} (r) > B^{2}(r,0)/2\mu $, otherwise the pressure of the external plasma is insufficient to confine the field that opens up and becomes more and more inclined as the external pressure decreases towards the photosphere. In that region, we can no longer neglect the effects of the tension force, so our simple model becomes invalid. However, the contribution of those surface layers to the perturbation of the stellar quadrupole moment is very small because of their relatively low density. Therefore, we apply our model from the base of the convection zone $r_{\rm b}$ up to some limit radius $r_{\rm L} < R$ where $\beta \equiv 2\mu p_{\rm e}/B^{2} = \beta _{\rm L}$ with the limiting parameter $\beta_{\rm L} $ arbitrarily fixed at $\beta_{\rm L} =3$.  

The perturbation of the density inside the magnetic flux tube can be computed by differentiating equation (\ref{pressure_balance}) with respect to the radial coordinate $r$ and taking into account that the pressure stratifications inside and outside the flux tube obey the equations 
\begin{equation}
\frac{\partial p_{\rm i,e}(r,\sigma)}{\partial r }= - \frac{GM(r) }{r^{2}} \rho_{\rm i, e}, 
\label{press_grad}
\end{equation}
where  $M(r)$ is the mass of the star inside the radius $r$, and $\rho$ the plasma density with the same meaning of the subscripts as in the case of the pressure. In this way, we find: 
\begin{equation}
\begin{array}{lll}
\rho_{\rm i}(r,\sigma) - \rho_{\rm e}(r) & = & {\displaystyle \frac{1}{2\mu} \frac{r^{2}}{GM(r)}\frac{\partial B^{2}(r, \sigma)}{\partial r} }\\
 & & \\
& = & {\displaystyle -\frac{2}{\mu} B_{0}^{2}(\sigma) r_{\rm b}^{4}  \frac{1}{GM(r)r^{3}} },
\label{density_pert}
\end{array}
\end{equation}
where we made use of Equation~(\ref{field_strength}) to compute the radial derivative of the field intensity. To compute the pressure gradients in equation~(\ref{press_grad}), we assume that the gravitational potential of the active star is spherically symmetric. This is a perfectly justified approximation given that the deviation of the local acceleration of gravity from the spherical symmetry does not exceed a few percents in most of the tidally distorted detached close binaries. 

\subsection{Perturbation of the gravitational quadrupole moment}
\label{quadrupole_pert}
The outer gravitational potential of the active star can be expressed as \citep[e.g.,][]{Applegate92}:
\begin{equation}
\Phi_{\rm G} = - \frac{GM}{r} - \frac{3G}{2r^{3}} \sum_{i,k} \frac{Q_{ik} x_{i} x_{k}}{r^{2}},
\end{equation}
where $M$ is the mass of the  star, $Q_{ik}$ its quadrupole moment tensor, and $x_{i}$ the Cartesian coordinates of a point outside the star in our reference frame as specified in Sect.~\ref{strong_fields} with $i,k =x, y, z$. The components of the quadrupole moment tensor can be expressed in terms of the components of the inertia tensor of the mass distribution of the star as:
\begin{equation}
Q_{ik} = I_{ik} -\frac{1}{3} \delta_{ik} {\rm Tr}\, I,
\end{equation}
where $\delta_{ik}$ is the Kronecker $\delta$ tensor, 
\begin{equation}
I_{ik} = \int_{V} \rho ({\bf x}) x_{i} x_{k}  \ dV,
\end{equation}
$\bf x$ being the position vector, and ${\rm Tr}\, I$  the trace of the inertia tensor, i.e., ${\rm Tr}\, I = I_{xx} + I_{yy} + I_{zz}$. 

The quadrupole moment due to stellar rotation and tidal deformation is considered steady in the reference frame of Figs.~\ref{model_overview} and~\ref{quadrupole_overview}, thus it does not contribute to the orbital period modulation in our model and can be neglected. We assume that only the density perturbation inside the radial flux tube $F$ produces a time-dependent contribution to the quadrupole moment  with an angle $\alpha$ in Figs.~\ref{model_overview} and~\ref{quadrupole_overview} that changes in time because we assume a small deviation of the stellar rotation from a perfect synchronization with the orbital motion (see below).  

To compute the components of the inertia tensor, it is useful to exploit the symmetry of the density configuration. To do so, we first consider the case when the axis of the flux tube $\hat{s}$ coincides with the axis $\hat{z}$ of our Cartesian frame and then apply a rotation to bring the flux tube in the equatorial plane as assumed by our model. We further assume that the flux tube has an angular radius $\theta_{0}$ (see Fig.~\ref{quadrupole_overview}) and indicate the density perturbation inside the flux tube as $\rho^{\prime} (r, \theta) = \rho_{\rm i} (r, \theta) - \rho_{\rm e}(r)$ as given by equation~(\ref{density_pert}) with $\sigma = r \sin \theta$ when $\hat{s} \equiv \hat{z}$. The perturbation  $\delta I_{zz}$ of the $I_{zz}$ component of the inertia tensor is:
\begin{equation}
\delta I_{zz} = \int_{V} \rho^{\prime} z^{2} \, dV. 
\end{equation}
Performing the integration in spherical coordinates with $z = r \cos \theta$ and making use of equation~(\ref{density_pert}), we find
\begin{equation}
\delta I_{zz} = -\frac{4\pi}{3} \frac{B_{0}^{2}}{\mu} r_{\rm b}^{4} (1-\cos^{3} \theta_{0}) {\cal J},  
\label{delta_i_zz}
\end{equation}
where ${\cal J}$ is the integral 
\begin{equation}
{\cal J} \equiv \int_{r_{\rm b}}^{r_{\rm L}} \frac{r^{\prime}}{GM(r^{\prime})} \, dr^{\prime}.  
\end{equation}
The perturbations of the principal components of the inertia tensor along the $\hat{x}$ and $\hat{y}$ axes in the equatorial plane are equal by symmetry. Since $x^{2} + y^{2} = r^{2} - z^{2}$, we can easily compute their sum and then find
\begin{equation}
\delta I_{xx} = \delta I_{yy} = - 2 \pi \frac{B_{0}^{2}}{\mu} \left( \frac{2}{3} -\cos \theta_{0} + \frac{1}{3} \cos^{3} \theta_{0} \right) r_{\rm b}^{4} {\cal J}.
\label{delta_i_xx}
\end{equation} 
 The perturbation of the trace of the inertia tensor is
\begin{equation}
\delta\, {\rm Tr} I = -4 \pi \frac{B_{0}^{2}}{\mu} (1- \cos \theta_{0}) r_{\rm b}^{4} {\cal J}. 
\label{delta_tr_i}
\end{equation}
The perturbations of the non-diagonal components of the inertia tensor are zero by symmetry: $\delta I_{xy} = \delta I_{xz} = \delta I_{yz} = 0$.

To compute the perturbations in the case of a flux tube the axis $\hat{s}$ of which is not along the polar axis of the star, we can apply two consecutive rotations of the reference frame, for example,  first around the $\hat{z}$ axis and  then around the transformed $\hat{y}$ axis, to bring the $\hat{z}$ axis to coincide with the $\hat{s}$ axis.  For the sake of simplicity, we assume that the $\hat{s}$ axis is in  the $xz$ plane and makes an angle $\zeta$ with the $\hat{z}$ axis, i.e., the colatitude of the flux tube is $\zeta$. In this case, we need only a rotation of an angle $\zeta$ around the $\hat{y}$ axis to find the perturbations of the components of the inertia tensor, that is:
\begin{equation}
\left\{
\begin{array}{lll}
\delta I_{x'x'} & = & \delta I_{xx} \cos^{2} \zeta + \delta I_{zz} \sin^{2} \zeta \\
\delta I_{y'y'} & = & \delta I_{yy} \\
\delta I_{z'z'} & = & \delta I_{xx} \sin^{2} \zeta +\delta I_{zz} \cos^{2} \zeta \\
\delta I_{x'y'} & = & 0 \\
\delta I_{x'z'} & = & (\delta I_{zz} - \delta I_{xx}) \sin \zeta \cos \zeta \\
\delta I_{y'z'} & = & 0 
\end{array}
\right.
\label{transf_inertia}
\end{equation} 
When $\zeta = \pi/2$, equations (\ref{transf_inertia}) provide the perturbations of the components of the inertia tensor due to a radial magnetic flux tube in the equatorial plane with $\hat{s} = \hat{x}$. 

Finally, the perturbations of the components of the quadrupole moment tensor are:
\begin{equation}
\delta Q_{ik} = \delta I_{ik} - \frac{1}{3} \delta_{ik}\left( \delta {\rm Tr} I \right). 
\end{equation}

\subsection{Equations of motion of the binary system}
\label{eqs_of_motion}

The equations of motion of a binary system when the gravitational field of one of the components is not axially symmetric have been investigated in the context of the rotation of Mercury and of some satellites of the solar system planets \citep{Goldreich66,GoldreichPeale66,MurrayDermott99}. \citet{Applegate89} made an application to close stellar binary systems that paved the way for the model presented in this paper. We shall neglect the torques due to tides and  stellar winds  because their timescales are much longer than the typical period of the orbital period modulation (cf. Sect.~\ref{tides_winds}) and consider only the effect of the non-axisymmetric perturbation of the gravitational quadrupole moment of the active star. 

The Lagrangian function ${\cal L}$ for our binary system can be written as
\begin{equation}
{\cal L} = {\cal T} - \Psi_{\rm G},
\end{equation}
where ${\cal T}$ is the kinetic energy of  the orbital motion and the rotation of the active star because the companion is treated as a point mass, while $\Psi_{\rm G}$ is the gravitational potential energy that includes  the term arising from the non-axisymmetric perturbation of the gravitational quadrupole moment. The expression of the kinetic energy when the spin and orbital angular momenta are aligned is  
\begin{equation}
{\cal T} = \frac{1}{2} m (\dot{r}^{2} + r^{2} \dot{f}^{2}) + \frac{1}{2} I_{\rm p} \dot{\varphi}^{2}, 
\end{equation}
where $m$  is the reduced mass of the system, $f$ the true anomaly of the orbital motion, $I_{\rm p}$ the moment of inertia of the active star about the $\hat{z}$ axis, that is,  $I_{\rm p} = I_{xx} + I_{yy}$, and $\varphi$ the angle of rotation of the active star around its spin axis (the $\hat{z}$ axis).  Note that both $\varphi$ and $f$ are given in an inertial reference frame, that is, they are measured with respect to a direction fixed in the inertial space and not with respect to the line joining the centres of the two components that is rotating in such a space with angular velocity $\dot{f}$.  On the other hand, the azimuthal coordinate $\phi$ is measured with respect to the orbit radius vector $\hat{x}$ joining the centres of the two components that is rotating in the inertial reference frame (cf. Sect.~\ref{strong_fields}). 

Because the axes $\hat{x}, \hat{y}$, and $\hat{z}$ are principal axes of inertia for the perturbed star, the expression of the gravitational potential energy of the system is
\begin{equation}
\begin{array}{lll}
\Psi_{G} & = & {\displaystyle -\frac{GMm_{\rm S}}{r} - \frac{3Gm_{\rm S}}{2 r^{3}} \left(\delta Q_{xx} \sin^{2} \theta \cos^{2} \alpha \right. } \\
 &  & {\displaystyle  \left. + \delta Q_{yy} \sin^{2} \theta \sin^{2} \alpha + \delta Q_{zz} \cos^{2} \theta \right) },
\label{potential_energy}
\end{array}
\end{equation} 
where $\delta Q_{xx}$, $\delta Q_{yy}$ and $\delta Q_{zz}$ are the perturbations of the components of the gravitational quadrupole moment computed as specified in Sect.~\ref{quadrupole_pert}, $\theta$ is the colatitude measured from the spin axis of the active star, and $\alpha \equiv  f-\varphi $ is the angle between the line joining the centres of the two stars and the axis of the magnetic flux tube  in the equatorial plane (see Fig.~\ref{model_overview}). For simplicity sake, we dropped the steady components of the quadrupole moment tensor due to the tidal and rotational deformations that do not contribute to the orbital period variation in the present model. 

Since the tensor $Q_{ik}$ is traceless, we define two scalars $\delta Q$ and $T$ such that
\begin{equation}
\delta Q_{xx} = \delta Q +T/2, \; \; \; \delta Q_{yy} = \delta Q -T/2, \mbox{ and } \delta Q_{zz} = -2 \delta Q, 
\end{equation}
with $T = \delta Q_{xx} - \delta Q_{yy}$. Substituting into equation~(\ref{potential_energy}), we find:
\begin{equation}
\Psi_{\rm G} = - \frac{GMm_{\rm S}}{r} - \frac{3Gm_{\rm S}}{2r^{3}} \left(-2 \delta Q + 3 \delta Q \sin^{2}\theta + \frac{1}{2} T \sin^{2}\theta \cos 2\alpha \right). 
\label{potential_energy_1}
\end{equation}
Assuming that the orbit lies on the equatorial plane ($\theta = \pi/2$), we derive the following equations of motion using the Lagrangian formalism:
\begin{equation}
\begin{array}{l}
{\displaystyle \ddot{r} - r \dot{f}^{2} + \frac{GM_{\rm T}}{r^{2}} + \frac{9GM_{\rm T}}{2r^{4}M} \left( \delta Q + \frac{1}{2} T \cos  2 \alpha \right) = 0, }\\
~\\
{\displaystyle I_{\rm p}  \ddot{\varphi} - \frac{3Gm_{\rm S} T}{2r^{3}} \sin 2 \alpha = 0, } \\
~\\
{\displaystyle m r^{2}  \ddot{f} + 2 m  r \dot{r} \dot{f} + \frac{3Gm_{\rm S} T}{2 r^{3}} \sin 2 \alpha = 0. } 
\end{array}
\label{eqs_motion}
\end{equation}
The effect of tides is that of making the orbit circular ($\dot{r} = 0$) and to align the spin and orbital angular momenta. The quadrupole moment variation does not excite any orbital eccentricity because the period of the oscillation of the angle $\alpha$ is much longer than the orbital period \citep[cf.][]{Phinney92,LanzaRodono01}, so we shall assume $\dot{r}=0$ in our equations of motion. 

The first of equations~(\ref{eqs_motion}) with $\ddot{r} = 0$ provides a generalization of the Kepler III law that we write
\begin{equation}
r^{3}\dot{f}^2 = G M_{\rm T} \left[1 + \frac{9}{2} \frac{1}{Mr^{2}} (\delta Q + \frac{1}{2} T \cos 2\alpha) \right] \simeq G M_{\rm T},
\label{kepler3}
\end{equation}
where we neglect the second term in the square brackets because it is of order of $10^{-6}$ or smaller with respect to the unity.

Now consider the second and the third of equations (\ref{eqs_motion}). Their sum can be immediately integrated with respect to the time to give the conservation of the total (orbital $+$ spin) angular momentum $J$ of the system:
\begin{equation}
I_{\rm p} \dot{\varphi} + m r^{2} \dot{f} = J, 
\label{ang_mom}
\end{equation}
Making use of the Kepler III law, we can recast this as
\begin{equation}
I_{\rm p} \dot{\varphi} + m (GM_{\rm T})^{2/3} \dot{f}^{-1/3} = J 
\label{ang_mom_1}
\end{equation}

By subtracting the second of the equations (\ref{eqs_motion}) from the third, considering that $\dot{r}=0$,  and  the definition of $\alpha \equiv f - \varphi$, we obtain an equation for $\alpha$ 
\begin{equation}
\ddot{\alpha} + \frac{1}{2}\omega_{\rm P}^{2} \sin  2\alpha = 0, 
\label{pendulum_eq}
\end{equation}
that is the equation of motion of a simple pendulum making oscillations of finite amplitude with  
\begin{equation}
\omega_{\rm P}^{2} = 3 \frac{Gm_{\rm S} T}{r^{3}}\left( \frac{1}{mr^{2}} + \frac{1}{I_{\rm p}} \right).  
\label{libr_freq}
\end{equation}
We  recast the expression for $\omega_{\rm p}$ using Kepler III law and introducing the mean orbital motion $n=2\pi/P$ as
\begin{equation}
\omega_{\rm p} = \sqrt{3} n \left( \frac{m_{\rm S}}{M_{\rm T}}\right)^{1/2} \left( \frac{T}{I_{\rm p}}\right)^{1/2} \left( 1 + \frac{I_{\rm p}}{m r^{2}} \right)^{1/2}.
\label{libr_freq_1}
 \end{equation}
Equation~(\ref{pendulum_eq}) has the first integral
\begin{equation}
\frac{1}{2} \dot{\alpha}^{2} + \frac{1}{2} \omega_{\rm P}^{2} \sin^{2} \alpha = \frac{1}{2} E^{2}, 
\label{first_integ}
\end{equation}
where $E \geq 0$ is a constant of the motion depending on the initial conditions. 
The positions of equilibrium falling at $\alpha = \pm \, k \pi$ with $k \in \mathbb{N}$ correspond to $E = 0$. The solutions of equation (\ref{first_integ}) require $E \geq \omega_{\rm p} \sin \alpha$ because $\dot{\alpha}^{2} \geq 0$. 
For $E \leq \omega_{\rm P}$, the angle $\alpha$ librates around a position of equilibrium making oscillations with amplitude $\alpha_{0} = \arcsin (E/\omega_{\rm P})$ with $\dot{\alpha} = 0$ when $\alpha = \pm \alpha_{0}$. On the other hand, if $E > \omega_{\rm P}$, the angle $\alpha$ circulates, that is,  it varies in a monotone way because $\dot{\alpha}$ is never equal to zero and never changes its sign. 

\subsubsection{Libration}

The period of  libration is given by:
\begin{equation}
P_{\rm libr} = \frac{4}{\omega_{\rm p}} K \left(\sin \alpha_{0} \right),
\label{libr_period}
\end{equation}
where $K(\gamma)$ with $\gamma < 1$ is the complete elliptical integral of the first kind (see Appendix~\ref{app1}). The period diverges for $E/ \omega_{\rm p} = \sin \alpha_{0}\rightarrow 1$ because 
$K(\gamma) \rightarrow \infty$ as $\gamma \rightarrow 1$. Note that the cycle of the orbital period is one half of the libration period because of the $2 \alpha$ argument in the equations of motion (\ref{eqs_motion}). The maximum and minimum of $\dot{\alpha}$ are given by:
\begin{equation}
\begin{array}{l}
 \dot{\alpha}_{\max}  = \omega_{\rm p} \sin \alpha_{0}  \\
  \dot{\alpha}_{\min}  =  -\omega_{\rm p} \sin \alpha_{0}  
\end{array}
\end{equation}
and corresponds to the extrema of the orbital period. These equations imply
\begin{equation}
\dot{\alpha}_{\max} - \dot{\alpha}_{\min} = 2 \omega_{\rm p} \sin \alpha_{0}.
\label{alpha_diff_libr}
\end{equation}

Twice during a libration period, $\dot{\alpha} = 0$ that corresponds to $\dot{\varphi} = \dot{f} \equiv \dot{f}_{0}$. This allows us to write the total angular momentum as
\begin{equation}
J = \left(I_{\rm p} + m r^{2}_{0} \right) \dot{f}_{0}
\label{jlibr}
\end{equation}
where $r_{0}$ is the orbital radius that corresponds to the orbital angular velocity $\dot{f}_{0}$. Using the conservation of the total angular momentum, the expression for $\dot{\alpha}$ can be written as
\begin{equation}
\dot{\alpha} = \dot{f} -\dot{\varphi} =  - \frac{J}{I_{\rm p}} + \left(1+\frac{m r^{2}}{I_{\rm p}}\right) \dot{f},
\end{equation}
where we applied equation (\ref{ang_mom}) to express $\dot{\varphi}$ in terms of $\dot{f}$. Making use of equation (\ref{jlibr}), Kepler III law to express $r$ in terms of $\dot{f}$ in the case of a circular orbit, and taking into account that the variation of $\dot{f}$ is very small in comparison with its mean value, we find
\begin{equation}
\dot{\alpha} = \left( 1 - \frac{m r_{0}^{2}}{3I_{\rm p}} \right) \left( \dot{f} - \dot{f}_{0} \right). 
\label{alpha_libr}
\end{equation}
In the case of a circular orbit, $\dot{f} = 2\pi/P$, therefore, we can use equation (\ref{alpha_libr}) to recast equation (\ref{alpha_diff_libr}) in terms of the relative variation of the orbital period $\Delta P / P$ introducing the mean orbital motion $n \equiv 2\pi/P$, where $P$ is the mean orbital period
\begin{equation}
\left(\frac{m r_{0}^{2}}{3 I_{\rm p}} -1 \right) \frac{\Delta P}{P} = 2 \left( \frac{\omega_{\rm p}}{n} \right) \sin \alpha_{0}. 
\label{alpha0_libr}
\end{equation}
This equation can be used to evaluate $\sin \alpha_{0}$ from the observations when $\omega_{\rm p}$ is determined from the quadrupole term $T$ that in turn depends on the magnetic field  of the flux tube considered in our model. The value of $\sin \alpha_{0}$ must be consistent with that derived from the length of the modulation cycle by means of equation (\ref{libr_period}). In practice, since $T$ is unknown, we iterate between equations (\ref{alpha0_libr}) and (\ref{libr_period}) until we find the values of $T$ and $\sin \alpha_{0}$ that satisfy both the two equations for the observed values of $P_{\rm mod} = P_{\rm libr}/2$ and $\Delta P / P$. For $r_{0}$ we take the orbital radius corresponding to the mean period $P$ given the very small variation of the period itself. 

Finally, we  compute the total variation of the angle $\alpha$ during one cycle of the orbital period modulation that must be close to $ 2 \alpha_{0} \leq 2\pi$ rad for consistency with equation~(\ref{alpha0_libr}). Considering the conservation of the total angular momentum, we find
\begin{equation}
| \Delta \alpha | = 2 \pi \left( \frac{mr_{0}^{2}}{3I_{\rm p}} -1 \right) \frac{|O-C|}{P},
\label{delta_alpha_libr}
\end{equation}
where $|O-C|$ is the amplitude of the difference between the observed mid-eclipse times $O$ and those computed with a constant-period ephemeris $C$. 

\subsubsection{Circulation}
\label{circ_theory}

When $E > \omega_{\rm P}$, the angle $\alpha$ circulates with the period  (cf. Appendix~\ref{app1}):
\begin{equation}
P_{\rm circ} = \frac{4}{\omega_{\rm p}} \left( \frac{\omega_{\rm p}}{E} \right) K\left( \frac{\omega_{\rm p}}{E} \right), 
\label{p_circ}
\end{equation}
that again diverges for $\omega_{\rm p}/E \rightarrow 1$. 
From the energy integral (\ref{first_integ}), we derive the minimum and the maximum of $\dot{\alpha}$ as
\begin{equation}
\begin{array}{l}
\dot{\alpha}_{\max}^{2} = E^{2} \\
\dot{\alpha}_{\min}^{2} = E^{2} -\omega_{\rm p}^{2}. 
\label{alpha_circ}
\end{array}
\end{equation}
Subtracting these equations from each other and with little algebra, we obtain
\begin{equation}
\left( \dot{\alpha}_{\max} + \dot{\alpha}_{\min} \right) \left(  \dot{\alpha}_{\max} - \dot{\alpha}_{\min} \right) = \omega_{\rm p}^{2}.
\label{eq42}
\end{equation}
Since the variation of $\dot{\alpha}$ is very small in comparison with its mean value, we can express $\dot{\alpha}_{\max}$ and $\dot{\alpha}_{\min}$ as
\begin{equation}
\begin{array}{l}
\dot{\alpha_{\max}} = \langle \dot{\alpha} \rangle + d\dot{\alpha,} \\
\dot{\alpha_{\min}} = \langle \dot{\alpha} \rangle - d\dot{\alpha}, 
\end{array}
\end{equation}
where $\langle \dot{\alpha} \rangle = (1/2) \left( \dot{\alpha}_{\max} + \dot{\alpha}_{\min} \right) $ is the mean value of $\dot{\alpha}$ along one cycle of the period modulation and 
$d \dot{\alpha}$ can be derived by differentiating equation~(\ref{ang_mom_1}) and applying the definition of the angle $\alpha$ as
\begin{equation}
d \dot{\alpha} = \left ( 1 - \frac{m r_{0}^{2}}{3I_{\rm p}} \right) d \dot{f},
\end{equation}
where $r_{0}$ is the orbital radius corresponding to the mean orbital period $P$ and $d \dot{f}$ is the difference of $\dot{f}$ with respect to the value $2\pi/P$ corresponding to the mean orbital period. 

We can introduce the mean degree of asynchronism of the rotation of the active component with respect to the mean orbital motion as
\begin{equation}
\eta_{\rm AS} \equiv \frac{n-\langle \dot{\varphi} \rangle }{n},
\end{equation}
where $n = 2\pi/P$ and $\langle \dot{\varphi} \rangle$ is the mean spin angular velocity of the active star along one cycle of the period modulation. The mean asynchronism $\eta_{\rm AS} >0$ because tides transfer angular momentum from the orbit to the rotation of the active primary to compensate for the angular momentum lost through its magnetized wind on timescales much longer than those of the orbital period modulation (cf. Sect.~\ref{tides_winds}). In defining $\eta_{\rm AS}$ we implicitly assumed that the active star is rotating rigidly with a mean angular velocity $\Omega = \langle \dot{\varphi} \rangle$. We shall discuss the validity of this hypothesis later. We express the mean of $\dot{\alpha}$ in terms of $\eta_{\rm AS}$ as
\begin{equation}
\langle \dot{\alpha} \rangle = n \eta_{\rm AS}. 
\end{equation}
Differentiating $\dot{f} = 2\pi/P$ and using the above definitions, finally we write equation~(\ref{eq42}) as 
\begin{equation}
2 \eta_{\rm AS} \left( \frac{m r_{0}^{2}}{3I_{\rm p}} -1 \right) \frac{\Delta P}{P} = \left( \frac{\omega_{\rm p}}{n}\right)^{2}
\label{eta_AS}
\end{equation} 
that can be used to evaluate $\eta_{\rm AS}$ from the observed amplitude of the orbital period modulation when the value of $\omega_{\rm p}$, that depends on $T$ (cf. equation~\ref{libr_freq}), is known. Another equation to compute the pendulum energy $E$ can be obtained from the second of equations~(\ref{alpha_circ}) giving
\begin{equation}
\left( \frac{E}{\omega_{\rm p}} \right)^{2} = 1 +\eta_{\rm AS} \left[ \eta_{\rm AS} - \left( \frac{mr_{0}^{2}}{3I_{\rm p}} -1 \right) \frac{\Delta P}{P} \right] \left( \frac{\omega_{\rm p}}{n} \right)^{-2}
\label{E_circ}
\end{equation}
As in the case of the librating solution, the value of the quadrupole moment $T$ is in general unknown, but we can iterate to find a consistent solution to the three equations~(\ref{p_circ}), (\ref{eta_AS}), and (\ref{E_circ}) that reproduces the observed period of the modulation $P_{\rm mod} = P_{\rm circ}/2$ and its amplitude $\Delta P /P$. 

\subsection{Tidal and stellar wind torques}
\label{tides_winds}

Now we add the effects of the tidal and wind torques that were previously neglected in equation (\ref{pendulum_eq})
\begin{equation}
\ddot{\alpha}+\frac{1}{2} \omega_{\rm p}^{2} \sin 2\alpha + \frac{\dot{\alpha}}{t_{\rm syn}} = - \frac{N_{\rm w}}{I_{\rm p}}, 
\label{rot_eq}
\end{equation}
where $t_{\rm syn}$ is the tidal synchronization timescale that depends on the tidal torque $\Gamma_{\rm tide}$ as $t_{\rm syn} \equiv I_{\rm p}/\Gamma_{\rm tide}$, and $N_{\rm w}$ is the torque of the magnetized stellar wind with $N_{\rm w} < 0$ in order to have a steady loss of angular momentum. 

By averaging equation (\ref{rot_eq}) over  the period of the modulation $P_{\rm mod}$, the only surviving terms are:
\begin{equation}
\frac{\langle \dot{\alpha} \rangle}{t_{\rm syn}} = -\frac{N_{\rm w}}{I_{\rm p}}. 
\label{tidal_balance}
\end{equation}
In other words, on timescales equal to $P_{\rm mod}$ or longer, the tidal torque extracts  angular momentum from the orbital motion of the binary to compensate for the loss in the active component due to its magnetized wind. To make this transfer possible, the rotation of the active component cannot be perfectly synchronized with the orbital motion, in other words, $\langle \dot{\alpha} \rangle \not= 0 $. Noting that $\langle \dot{\alpha} \rangle = n - \Omega$, we define the expected degree of asynchronism of  the active component $\eta_{\rm wt}$ as 
\begin{equation}
\eta_{\rm wt} = \frac{n-\Omega}{n} = - \frac{N_{\rm w} t_{\rm syn}}{n I_{\rm p}},
\label{tidal_balance1}
\end{equation}
where we made use of equation~(\ref{tidal_balance}). 

To evaluate the tidal torque $\Gamma_{\rm tide}$ acting on the active star, we follow  \citet{MardlingLin02} considering a circular orbit (cf. their equation 54):
\begin{equation}
\Gamma_{\rm tide} = - \frac{9}{2Q^{\prime}} \frac{m_{\rm S}^{2}}{M_{\rm T}} \sqrt{G M_{\rm T} R} \left( \frac{R}{a} \right)^{9/2} \left(\Omega - n \right), 
\end{equation}
where $Q^{\prime}$ is the modified tidal quality factor of the active component that describes the efficiency of the tidal energy dissipation inside the star. 
Therefore, the tidal timescale in equation~(\ref{rot_eq}) is given by 
\begin{equation}
t_{\rm syn} = \frac{2 Q^{\prime}}{9} \frac{M_{\rm T}}{m_{\rm S}^{2}}  \frac{I_{\rm p}}{\sqrt{GM_{\rm T} R}} \left( \frac{a}{R} \right)^{9/2}. 
\label{tsyn}
\end{equation}
The lag angle $\alpha_{\rm tide}$ between the tidal bulge and the line joining the centres of the two components is related to $Q^{\prime}$ as 
\begin{equation}
\alpha_{\rm tide} \sim \frac{1}{Q^{\prime}}, 
\end{equation}
where $Q^{\prime} \gg 1$.  The modified tidal quality factor depends on the internal structure of the star and on the tidal frequency, $\hat{\omega} = 2( \Omega -n)$, because the semidiurnal tide dominates in the case of a circular and coplanar orbit \citep{Ogilvie14}. For nearly synchronous close binaries ($| \hat{\omega} | < 2 \Omega$), the excitation of inertial waves by the time-varying tidal potential increases the dissipation in a remarkable way, so we  assume $Q^{\prime} \sim 10^{5}-10^{6}$ corresponding to a strong tidal coupling between the components, further increased by the fast rotation of the  stars \citep[cf.][who found $Q^{\prime} \propto \Omega^{-2}$]{OgilvieLin07}. On the other hand, the tidal dissipation is much lower in the case of the stars hosting hot Jupiters because they are far from synchronous rotation ($| \hat{\omega}| \geq 2 \Omega$). In that case $Q^{\prime} \sim 10^{7}-10^{8}$ \citep{OgilvieLin07,Bonomoetal17,CollierCameronJardine18}.

The large values of $Q^{\prime}$ imply  very small values of $\alpha_{\rm tide}$ which means that the non-axisymmetric tidal bulge is almost perfectly aligned with the two components. Note that $\alpha_{\rm tide}$ cannot oscillates because the tidal bulge of the active star is always lagging the orbital motion of the companion to transfer angular momentum from the orbit to the stellar rotation in order to compensate for the wind torque.  
Therefore, we neglect the fixed quadrupole component associated with the tidal bulge in equation (\ref{potential_energy}). 

The wind torque $N_{\rm w}$ can be obtained from equation~(4) of \citet{Amardetal16} considering that the angular momentum loss rate is in the saturated regime in the case of the fast-rotating active components of close binary systems. On the other hand, in the case of stars with hot Jupiters, we can use their equation~(5) because the stellar wind is in the unsaturated regime, except for young hosts rotating faster than the saturation angular velocity $\Omega_{\rm sat} \simeq (10-12) \Omega_{\odot}$. The expression for $N_{\rm w}$ in the saturated regime is:
\begin{equation}
N_{\rm w} = - K_{\rm w} \, \Omega\, \Omega_{\rm sat}^{2} \, \left( \frac{R}{R_{\odot}} \right)^{1/2} \, \left( \frac{M}{M_{\odot}} \right)^{-1/2},
\label{wind_torque}
\end{equation}
with $K_{\rm w} = 2.7 \times 10^{40}$ kg~m$^{2}$~s. 
Considering a typical RS CVn system with an active component of $M=1.3$~M$_{\odot}$ and $R= 4.0$~R$_{\odot}$, a secondary with $m_{\rm S} = 1.3$~M$_{\odot}$, $a = 4 R$, $P = 3$~d, and $Q^{\prime} = 10^{6}$, we obtain $\eta_{\rm wt} = 3.4 \times 10^{-6}$ from equation~(\ref{tidal_balance1}) that is  compatible with the values of $\eta_{\rm AS}$ that will be derived from our model in Sect.~\ref{applications}. Note that a stronger tidal interaction as parameterized by a smaller $Q^{\prime}$ would imply a smaller value of $\eta_{\rm AS}$ because a smaller degree of asynchronism would be sufficient to drain enough angular momentum from the orbit to balance the losses in the wind. 

When comparing the present estimates with the results obtained with the model in Sect.~\ref{circ_theory}, we do not expect that $\eta_{\rm AS}$ and $\eta_{\rm wt}$ will coincide because of the large uncertainties in the tidal and wind theories. The expressions for $Q^{\prime}$ and $N_{\rm w}$ have been calibrated with main-sequence F, G, and K stars belonging to open clusters of different ages and with rotation periods  $\ga 1-2$ days. Therefore, their extrapolations to the case of active subgiant stars in Algols and RS CVn's systems or to the K or M main-sequence stars in CVs or PCEBs with rotation periods below 0.5 days can be remarkable in error. In other words, we cannot expect a coincidence between $\eta_{\rm AS}$ and $\eta_{\rm wt}$ better than within $1-2$ orders of magnitude.

\subsection{Variation of stellar rotation}
\label{star_rotation}
Combining the conservation of the total angular momentum (equation~\ref{ang_mom}) with Kepler III law and considering that the system is nearly synchronous ($\Omega \sim n$), we can compute the variation of the stellar angular velocity $ \Omega$ associated with a relative variation of the orbital period $\Delta P/P$:
\begin{equation}
 \frac{\Delta \Omega}{\Omega} = - \frac{mr_{0}^{2}}{3 I_{\rm p}} \frac{\Delta P}{P}. 
 \label{rotation_change}
\end{equation}
Therefore, our model predicts that a decrease of the orbital period $\Delta P / P < 0$ is accompanied by a spin up of the active component, while an increase of the period is associated with a spin down of the stellar rotation. For typical values $\Delta P /P \sim (0.1-1) \times 10^{-5}$ and $mr^{2}/I_{\rm p} \sim 25-100$, we predict variations with a relative amplitude of $\Delta \Omega / \Omega \sim (0.1-10) \times 10^{-4}$ that could be observable if the lifetimes of starspots and magnetic features, used as rotation tracers, were at least of the order of $10^{3}$ rotation periods, that is, comparable with $2\pi/\Delta \Omega$ if a variation $\Delta \Omega$ is to be measured.  The existence of such extremely long-lived features is questionable \citep{BradshowHartigan14,Gilesetal17}, therefore, such very small rotation  changes are likely impossible to measure. Note that these oscillations of the mean angular velocity of the active component are  averaged out when we consider timescales equal to or longer than the modulation cycle $P_{\rm mod}$. Thus they do not contribute to the mean level of asynchronous rotation $\eta_{\rm AS}$ as defined in Sect.~\ref{eqs_of_motion} or to $\eta_{\rm wt}$ in Sect.~\ref{tides_winds},  the amplitude of which is two or three orders of magnitude smaller. 

The torque accelerating or decelerating the stellar rotation basically acts on the magnetic flux tube where the density perturbation is localized, thus the transferred angular momentum needs  to be redistributed through the whole convection zone for the effects to be observable. Reynolds stresses produced by turbulent convective motions can be regarded as  relevant transporters of angular momentum producing this redistribution over a characteristic  timescale $\tau_{\rm RS} \sim R^{2}/\nu_{\rm turb}$, where $R$ is the radius of the star and $\nu_{\rm turb} = (1/3) l v_{\rm c}$ the kinematic turbulent diffusivity with $l$ being the mixing length and $v_{\rm c}$ the convective velocity \citep[cf.][]{RuedigerHollerbach04}. In rapidly rotating late-type stars, the turbulent transport becomes anisotropic, thus  transport coefficients should be substituted by tensors the components of which are strongly dependent on angular velocity and magnetic field components \citep[e.g.,][]{Warneckeetal18}. Generally, the transport of angular momentum is quenched with respect to the simple mixing-length estimate adopted above with a decrease of $\nu_{\rm turb}$ by approximately one order of magnitude when the magnetic field is $\sim 3$ times the equipartition value.  Nevertheless, in a strongly magnetized convection zone,  the Maxwell stresses  may redistribute the angular momentum on a timescale comparable with the Alfven crossing time $\tau_{\rm ALF}$ along the magnetic flux tube from its base at $r=r_{\rm b}$ to the surface at $r=R$, that is shorter than the turbulent diffusion time. Therefore, the actual redistribution time in our active stars should fall in between the two extremes $\tau_{\rm ALF}$ and $\tau_{\rm RS}$ depending on the relative contributions of the two mechanisms that are not possible to predict without a detailed modelling. 

As we shall see in Sect.~\ref{applications}, in the active components of close binary systems, neglecting turbulent transport quenching, $\tau_{\rm RS}$ is comparable or longer than the modulation cycle of the orbital period, while $\tau_{\rm ALF}$ is remarkably shorter, suggesting that a significant fraction of the transferred angular momentum can be redistributed over the whole convection zone of the active star during the modulation cycle thanks to the Maxwell stresses. In other words, the above estimate of $\Delta \Omega/ \Omega$ can be regarded as an upper limit to the observable variation, although the latter should not be much smaller. 

A relative variation of the rotation $\la10^{-3}$ is comparable with or smaller than the observed amplitudes of the differential rotation in the active components of RS CVn binaries or single rapidly rotating solar-like stars such as AB Dor that has a rotation period of 12 hrs. The strong internal magnetic fields of such stars produce fluctuations of the surface differential rotation that have amplitudes comparable with the differential rotation itself owing to the effects of the Lorentz force  \citep[e.g.,][]{DonatiCollierCameron97,Donati99,CollierCameronDonati02,CollierCameronetal02,Donatietal03,Barnesetal05}. Therefore,  the observation of the cyclic variations predicted by our model may be hampered by the internal redistribution of the angular momentum in the active component due to its strong magnetic fields. 

\subsection{Systems with hot Jupiters}
\label{HJs}

Stars hosting hot Jupiters are generally far from synchronous rotation, except in a few notable cases, such as that of $\tau$~Bootis \citep{DamianiLanza15,Borsaetal15}. When $\eta_{\rm AS} \sim 1$, we can only have circulation and the expected orbital period modulation is very small according to equations (\ref{libr_freq}) and (\ref{eta_AS}), that is, of the order of $\Delta P/P \sim 10^{-10}$ even in the case of very active stars. The reason is that the exchange of angular momentum between the orbit and the spin of the active star becomes very inefficient because the angle $\alpha$ changes on the timescale of the stellar rotation, that is, much shorter than the typical period $P_{\rm mod}$ of the modulation cycles. Only if the system is close to synchronization, so that $\alpha$ changes slowly, the small torque produced by the density perturbation in the magnetized plasma has time to transfer enough angular momentum to produce an observable variation of the orbital period. 

We conclude that our model does not predict observable orbital period modulations in systems with hot Jupiters, unless the star is very active and the system close to synchronization. However, even if those requirements are fulfilled, we expect to see modulations with an amplitude smaller than in the case of active close binaries because the term $mr_{0}^{2}/I_{\rm p}$ is larger by a factor of $\sim 5-20$ in systems with hot Jupiters.

\section{Applications}
\label{applications}
In this section, we consider only three representative cases to illustrate the application of the model developed in Sect.~\ref{model_desc} and defer to a successive work a systematic application  to a larger sample of binary stars. 

The relevant parameters of our systems and of their active components are listed in Table~\ref{system_params}. This table reports from the left to the right, the name of the binary system; the mass $M$ of the active component; its radius $R$; its luminosity $L$; the relative radius at the base of its convection zone $r_{\rm b}/R$; the mass fraction $m_{\rm b}/M$ at radius $r_{\rm b}$; the moment of inertia of the active star about its spin axis $I_{\rm p}$; the mass $m_{\rm S}$ of the secondary component of the system; the orbital period $P$; the ratio of the orbital to the stellar moment of inertia $mr^{2}_{0}/I_{\rm p}$; the expected relative asynchronization $\eta_{\rm wt}$ according to the tidal and wind theory in Section~\ref{tides_winds}, computed for $Q^{\prime} = 10^{6}$ and assuming that the wind angular momentum loss rate is saturated; an estimate of the turbulent diffusion time $\tau_{\rm RS} \sim R^{2}/\nu_{\rm turb}$, where $\nu_{\rm turb}$ is the mean of the turbulent diffusivity (see Section~\ref{star_rotation}); and the literature reference for the system parameters.   We do not consider rotational or magnetic quenching of the turbulent diffusion, therefore, the estimated $\tau_{\rm RS}$ is a lower limit that can differ by $1-2$ orders of magnitude from the true timescale in the case of very active stars (cf. Section~\ref{star_rotation}). 

The parameters of the active components are derived from internal structure models computed for the given masses $M$ with the MESA \citep{Paxtonetal11} web interface assuming a metallicity $Z=0.02$, a ratio of the mixing length $l$ to the local pressure scale height $H_{\rm p}$, $l/H_{\rm p}=2.0$, and without including the structure effects of rotation and overshooting \footnote{http://mesa-web.asu.edu/}. The mass $M$ of the primary, the  orbital period $P$, the mass of the secondary $m_{\rm S}$, and the orbital radius $r_{0}$ come from the indicated literature references, respectively.  

The vertical magnetic flux tube in our model is assumed to be in the equatorial plane ($\zeta = \pi/2$) with an angular radius $\theta_{0}=30^{\circ}$ in all the considered active stars giving a filling factor of 6.7 percent, that is, perfectly compatible with the starspot coverage observed in RS CVn systems and other active binaries that ranges between $\approx 10$ and $\approx 50$ percent \citep[e.g.][]{Rodonoetal95,Neffetal95}. Note that increasing $\theta_{0}$ for a fixed value of the quadrupole moment variation decreases the magnetic field strength at the base of the flux tube $B_{0}$  (cf. equations~\ref{delta_i_zz}, \ref{delta_i_xx}, and~\ref{delta_tr_i}). Our choice of a relatively small filling factor at the base of the stellar convection zone, corresponding to the adopted $\theta_{0}=30^{\circ}$, is therefore rather conservative and gives an upper limit for the field strength in the flux tube.  

\begin{table*}
\caption{System and active star parameters. }
\begin{tabular}{ccccccccccccc}
\hline
System  & $M$ & $R$ & $L$ & $r_{\rm b}/R$ & $m_{\rm b}/M$ & $I_{\rm p}$ & $m_{\rm S}$ & $P$ & $mr_{0}^{2}/I_{\rm p}$ & $\eta_{\rm wt}$ & $\tau_{\rm RS}$ & Ref. \\
 & (M$_{\odot}$) & (R$_{\odot}$) & (L$_{\odot}$) & & &  (kg~m$^{2})$ & (M$_{\odot}$) & (days) & & & (yrs) & \\
 \hline
 HR 1099 & 1.3 & 4.255 & 9.327 & 0.193 & 0.257 & $3.218 \times 10^{48}$ & 1.05 & 2.83774 & 24.303 & $6.675 \times 10^{-7}$ & $\sim 56$ & 1 \\ 
 V471 Tau & 1.0 & 0.941 & 0.834 & 0.724 & 0.975 & $7.056 \times 10^{46}$ & 0.875 & 0.52118 & 67.14 & $2.966 \times 10^{-6}$ & $\sim 36$ & 2 \\
 NN Ser & 0.15 & 0.171 & $3.812 \times 10^{-3}$ & 0.0 & 0.0 & $8.981 \times 10^{44}$ & 0.57 & 0.13 & 112.06 &  $1.063 \times 10^{-3}$ & $\sim 1.8$ & 3 \\
 \hline
\end{tabular}
\begin{flushleft}
Note. References for system parameters: 1:~\citet{Lanzaetal06}; 2:~\citet{Vaccaroetal15}; 
3:~\citet{Brinkworthetal06}. 
\end{flushleft}
\label{system_params}
\end{table*}

\subsection{HR 1099}
HR 1099 is a detached binary of the RS CVn type for which the orbital period modulation is very large ($\Delta P/P \sim 9 \times 10^{-5}$) and cannot be explained by a light-time effect as confirmed by the constancy of the radial velocity of its barycentre \citep{Donati99,FrascaLanza05}.  The length of the cycle of its orbital modulation is $P_{\rm mod} \approx 36$~yrs, while the starspot cycle of the active component has a period of $\sim 14-19$~yrs \citep[e.g.][]{Lanzaetal06,Muneeretal10,Peldelwitzetal18}. The analysis of \citet{Muneeretal10} found a somewhat smaller relative amplitude of the period modulation $\Delta P / P \sim 5 \times 10^{-5}$, but we adopt the larger value given in previous studies to put a stronger constraint on our model.  Indeed, this system has been considered as a benchmark for a comparison of different models of orbital period modulation \citep[e.g.,][]{Lanza05,Lanza06,Volschowetal18}.

Considering first the case of a libration of the angle $\alpha$, we solve simultaneously equations~(\ref{libr_period}) and (\ref{alpha0_libr}) by successive iterations and find the model parameters listed in the first row of Table~\ref{table_libration}. In this table, we report from the left to the right the name of the binary system; the modulation period $P_{\rm mod} = P_{\rm libr}/2$ as derived from the model equations; the quadrupole moment $T$; the libration pulsation $\omega_{\rm p}$; the sine of the limiting angle of libration $\alpha_{0}$; the estimate of the maximum excursion of  the angle $\alpha$ based on the observed amplitude $O-C$ and equation~(\ref{delta_alpha_libr}); the magnetic field $B_{0}$ in the flux tube at the base of the convection zone; the total magnetic energy $E_{\rm mag}$ of the field in the flux tube; the time required by the stellar luminosity $L$ to supply the energy $E_{\rm mag}$; the radius $r_{\rm L}$ where the plasma pressure becomes smaller than three times the magnetic energy density; and the Alfven crossing time along the magnetic flux tube from the base of the convection zone to the surface of the star. 

The librating solution is acceptable in the case of HR~1099, although the energy of the oscillator is very close to the limit $E/\omega_{\rm p} = \sin \alpha_{0} = 1$ beyond which the angle $\alpha$ circulates. The value of $| \Delta \alpha| $ is uncertain in this system because we do not have observed a complete cycle of the modulation yet \citep[cf. ][]{Lanzaetal06}. Assuming the revised ephemeris of \citet{Muneeretal10},  just one complete cycle appears to have been covered with an $O-C$ amplitude of 0.17 days that yields $|\Delta \alpha | = 144^{\circ}.14$ giving $\sin \alpha_{0} = 0.9514$ that is not incompatible with the value found from the libration model, given the uncertainties. 

The magnetic field intensity and energy found in the case of libration are perfectly feasible and require a small amount of the energy available from the stellar luminosity to support the hydromagnetic dynamo that should maintain the field against turbulent diffusion, the latter operating with a timescale $\tau_{\rm RS}$ comparable or slightly longer than the modulation cycle. 

On the other hand,  the parameters of the circulation models are listed in Table~\ref{table_circulation}, where the contents of the columns are the same as in Table~\ref{table_libration}, except for $\omega_{\rm p}/E$ and the relative asynchronization $\eta_{\rm AS}$ that comes from equation~(\ref{eta_AS}). In the case of HR~1099, the circulating solution requires a ratio $\omega_{\rm p}/E$ extremely close to 1 because even a deviation as small as $10^{-8}$ is not sufficient to reproduce the period of the modulation (cf.  the first row of Table~\ref{table_circulation}).  Therefore, in the case of HR~1099, our preference is given to  the libration model. 

The change in stellar rotation along a cycle of the orbital period modulation, computed with equation~(\ref{rotation_change}),  is $\Delta \Omega / \Omega \simeq 6.94 \times 10^{-4}$. It is probably an upper limit for the observable variation because the turbulent diffusion of the angular momentum takes a timescale comparable or longer than the modulation cycle. On the other hand, if the angular momentum is mainly redistributed by the Maxwell stresses, the stellar rotation can be adjusted on a shorter timescale because $\tau_{\rm ALF}$ is $\sim 0.4$~yr.  

The kinetic energy variation associated with the transfer of angular momentum from the orbit to the spin of the active star is  $\Delta {\cal T} = 1.541 \times 10^{36}$~J. Even if all the kinetic energy $\Delta {\cal T}$ were dissipated inside the active component, the mechanism would still be perfectly feasible from an energetic point of view because the stellar luminosity requires only 13.6 yrs to supply this amount of energy, that is only 1/3 of the duration of the modulation cycle. As a matter of fact, the amount of dissipated energy is certainly lower than $\Delta {\cal T}$ because the shear associated with the differential rotation is very small in the active component of HR~1099 \citep[cf.][]{Donatietal03}. From an observational  point of view, the relative change in the rotation rate of the starspots along the activity cycle of HR 1099 is $\sim 1.1 \times 10^{-3}$ \citep{Lanzaetal06}, that is larger than the predicted variation of the stellar rotation due to the spin-orbit angular momentum exchanges. Therefore, the variation of the rotation rate of the starspots associated with their latitudinal migration along the cycle may hide the variation expected along the longer cycle of the orbital period modulation. 

\begin{table*}
\caption{Parameters of the libration models computed for the orbital period modulation of the listed binary systems (see text for discussion). }
\begin{tabular}{ccccccccccc}
\hline
System & $P_{\rm mod}$ & $|T| /I_{\rm p}$ & $\omega_{\rm p}/n$ & $\sin \alpha_{0}$ & $ |\Delta \alpha | $ & $B_{0}$ & $E_{\rm mag}$ & $E_{\rm mag}/L$ & $r_{\rm L}/R$ & $\tau_{\rm ALF}$ \\
 & (yr) & & & & (deg)  & (T) & (J) & (yr) & &  (yr) \\ 
 \hline 
 HR 1099 & 35.95 & $7.328 \times 10^{-8}$ & $3.198 \times 10^{-4}$ & 0.99926358 & 324.31 &  7.8124 & $3.064 \times 10^{33}$ & 0.027 & 0.99498 & 0.395 \\
 V471 Tau & 35.25 & $4.213 \times 10^{-10}$ & $2.451 \times 10^{-5}$ & 0.74368241 & 51.55  & 0.8400 & $ 6.931 \times 10^{30}$ & $6.82 \times 10^{-4}$ & 0.9987 & 0.078 \\
 NN Ser & 16.02 & $1.001 \times 10^{-10}$ & $1.548 \times 10^{-5}$ & 0.87638261 & 69.91 &  0.3550 & $4.444 \times 10^{27}$ & $9.57 \times 10^{-5}$ & 1.0 & 2.727 \\
 \hline 
\end{tabular}
\label{table_libration}
\end{table*}
\begin{table*}
\caption{Parameters of the circulation models computed for the orbital period modulation of the listed binary systems (see text for discussion).}
\begin{tabular}{ccccccccccc}
\hline
System & $P_{\rm mod}$ & $|T| /I_{\rm p}$ & $\omega_{\rm p}/n$ & $\omega_{\rm p}/E$ & $\eta_{\rm AS}$ & $B_{0}$ & $E_{\rm mag}$ & $E_{\rm mag}/L$ & $r_{\rm L}/R$ & $\tau_{\rm ALF}$ \\
 & (yr) & & & & & (T) & (J) & (yr) & &  (yr) \\ 
\hline
HR 1099 & 28.50 & $5.854 \times 10^{-7}$ & $ 9.038 \times 10^{-4}$ & $1-1.0 \times 10^{-8}$ & $6.391 \times 10^{-4}$ & 22.162 & $2.463 \times 10^{34}$ & 0.217 & 0.9901 & 0.139 \\
V471 Tau & 35.44 & $1.882 \times 10^{-9}$ & $5.180 \times 10^{-5}$ & $0.997558$ & $3.681 \times 10^{-5}$ & 1.779  & $3.101 \times 10^{31}$ & 0.003 & 0.9978 & 0.037 \\ 
NN Ser & 16.61 & $6.151 \times 10^{-10}$ & $3.839 \times 10^{-5}$ & 0.999897 & $2.715 \times 10^{-5}$ & 0.880 & $2.732 \times 10^{28}$ & $5.9 \times 10^{-4}$ & 1.0 & 1.100 \\
\hline
\end{tabular}
\label{table_circulation}
\end{table*}

\subsection{V471 Tauri}
The second system to which we apply our model is the PCEB V471~Tauri that was also considered in several previous investigations of the orbital period modulation. 
It shows a modulation with an amplitude $\Delta P /P \simeq 8.5 \times 10^{-7}$ and $P_{\rm mod} \sim 35$~yrs \citep[cf.][]{Marchionietal18}. We assume that the modulation is due to the mechanism proposed in this paper, although the possibility of a light-time effect due to a third body cannot be completely ruled out with the present data. 

Both the libration and the circulation models are feasible for this system from the point of view of the parameters and the energy required to support the magnetic field. Only a very small fraction of the stellar luminosity is required to power the hydromagnetic dynamo to maintain the magnetic field against turbulent diffusion that has a timescale comparable with the modulation cycle of the orbital period. In the case of  libration, the excursion $|\Delta \alpha |$ computed from the observed $O-C$ amplitude and period is smaller than $2\alpha_{0}$, but again this could be a consequence of having observed only one cycle of the modulation, so its period and amplitude are not well constrained \citep[cf.][]{Marchionietal18}. 

The relative amplitude of the variation of the angular velocity as given by equation~(\ref{rotation_change}) is $\Delta \Omega / \Omega = 3.816 \times 10^{-5}$ that is too small to be observable. The associated variation of the kinetic energy of the spin of the active star is $\Delta {\cal T} = 5.241 \times 10^{34}$~J with a luminosity timescale $\Delta {\cal T}/L = 5.16$ yrs, that is less than $1/4$ of  the modulation cycle. Therefore, the possible dissipation of the kinetic energy $\Delta {\cal T}$ inside the active component has no impact on the energetic feasibility of the mechanism.

\subsection{NN Serpentis}

This is a PCEB detacted binary, worth of investigation because the modulation of the mid-eclipse times, interpreted as a light-time effect, points to  the presence of two planetary companions around the binary whose orbits can be stable on a timescale comparable with the estimated age of the system \citep{Brinkworthetal06,Boursetal16}. The amplitude of the modulation is $\Delta P/ P \sim 7.5 \times 10^{-7}$ with $P_{\rm mod} \sim 16$~yrs  \citep{Boursetal16}. 

Here we explore an interpretation in terms of an intrinsic orbital period modulation due to the proposed model. The active component  is a very low-mass main-sequence star accompanied by a more massive  white dwarf. Assuming a mass of only $0.15$~M$_{\odot}$ for the active star, its internal structure is fully convective. To avoid a divergence of the magnetic field inside our model flux tube at the centre of the star, we assume that the flux tube extends from the mid of the convection zone to the surface, thus the value $B_{0}$ refers to a base radius $r_{\rm b} = 0.5\,R$. 

The libration model seems to be preferable for this system in terms of a less extreme value of $\omega_{\rm p}/E$, although again $|\Delta \alpha|$ is not coincident with $2 \alpha_{0}$. 
However, the relative variation of the stellar spin  coming from equation~(\ref{rotation_change}) is $\Delta \Omega / \Omega = 2.789 \times 10^{-5}$ implying a kinetic energy change of $\Delta {\cal T} = 7.837 \times 10^{33}$~J and a luminosity timescale of $\Delta {\cal T}/L = 168.75$ yrs, much longer than the modulation cycle. We conclude that, even if less than 10 percent of the kinetic energy $\Delta {\cal T}$ is dissipated during the operation of the mechanism, the very small stellar luminosity does not appear capable to supply the required energy along one modulation cycle. This gives support to the interpretation of the apparent orbital period changes in terms of a light-time effect. 

\section{Discussion and conclusions}
We have introduced a new model to explain the orbital period modulation observed in detached and semi-detached close binary systems with a late-type magnetically active component. An illustrative application to three representative  systems  shows that the model is capable of accounting for the observations, except in the case of an extremely low-mass active component star (cf. Sect.~\ref{applications}). 

Our model assumes that an internal magnetic field produces a non-axisymmetric quadrupole moment that persists for timescales longer than the orbital period modulation. This is  suggested by the observation of active longitudes that persist for very long times in active stars (see Sect.~\ref{model_overview_text}). From a theoretical point of view, we may invoke $\alpha^{2}$-type dynamos in those rapidly rotating stars, where the helicity of convective motions imparted by the Coriolis force may dominate over the effects of the differential rotation,  to produce steady magnetic fields \citep{Ruedigeretal02,RuedigerHollerbach04}. A mixture of steady and oscillating fields can be produced according to the profile of the hydromagnetic $\alpha$-effect or by the simultaneous operation of different kinds of dynamos in different layers of the stellar convection zone, thus accounting for the magnetic cycles observed in those stars \citep{Ruedigeretal02,Olahetal09}. Stellar cycles on century timescales can account for very long-term orbital period modulations such as those observed in the prototype Algol \citep{Soderhjelm80} because they can modulate the internal magnetic field and the associated quadrupole moment on those timescales.  

Our model is based on a coupling of the spin of the active star with the orbital motion of the binary that is directly mediated by the non-axisymmetric stellar quadrupole moment and not by tides whose timescale is much longer. The cyclic exchange of angular momentum between the stellar spin and the orbit is responsible for the modulation of the orbital period. A consequence of this exchange is the variation of the stellar rotation along the cycle of the modulation with relative amplitudes of the order of $10^{-5}-10^{-4}$ (cf. Sections~\ref{star_rotation} and~\ref{applications}). They are in general too small to be detectable, but may play a role in modulating stellar activity because they might excite torsional oscillations in the large-scale stellar poloidal field that could account for the approximately 2:1 ratio of the periods of the orbital modulations and starspot activity cycles as observed in some systems \citep[see][for details]{Lanzaetal98,LanzaRodono04}.   

The most attractive feature of the proposed model is its capability of easily verifying all the constraints imposed by energy conservation. In particular, the energy available from the stellar luminosity along one cycle of the modulation is more than sufficient to support the proposed mechanism, except in very low mass stars such as in NN~Ser ($M \la 0.15$~M$_{\odot}$). This is not the case with the quadrupole moment change models of \citet{Applegate92} or \citet{Lanzaetal98} as demonstrated by, e.g.,  \citet{Lanza06} or \citet{Volschowetal18}.

Our model gives observable orbital period modulations only in systems that are close to tidal synchronization, while predicting negligible orbital period variations in  transiting hot Jupiter systems because they are generally far from synchronization (cf. Section~\ref{HJs}). Indeed, the change in the angle $\alpha$ between the principal axis of inertia of the active component and the orbital radius vector produces an additional very small variation of the orbital period because $\alpha$ appears in the generalized Kepler III law in equation~(\ref{kepler3}). The relative variation  is $\Delta P/P \sim 10^{-8}-10^{-7}$ in RS CVn and Algol binaries and $\Delta P/P \sim 10^{-11}-10^{-10}$ in PCEBs and CVs systems having a main-sequence active star. Therefore, they are much smaller than the variations produced by the spin-orbit coupling considered in our model and can be completely neglected in the case of stellar binary systems. Nevertheless, they could be detectable in the case of hot Jupiter systems, if very accurate measurements of their orbital periods become available because  the expected relative changes are at the level of $10^{-11}-10^{-10}$. 

Finally, we consider the similarity between our model and that proposed by \citet{Applegate89}. Although the basic equations are similar, he considered much larger values of $T/I_{\rm p}$ leading to modulation periods remarkably shorter than the observed modulation cycles in close binaries. The estimates of the tidal dissipation efficiency available at that time were  remarkably smaller than what we used in this work and this may have prevented Applegate from discovering the kind of solutions exploited in the present model. In particular, the strong tidal coupling occurring in late-type synchronized binaries was not recognised until the works by \citet{MeibomMathieu05} and \citet{OgilvieLin07}. It represents a crucial ingredient of the present model because it leads to a small level of asynchronization $\eta_{\rm AS}$ that implies a slow variation of the angle $\alpha$. This gives sufficient time for the torque associated with the small non-axisymmetric quadrupole moment to transfer angular momentum back and forth from the stellar rotation to the orbit (cf. Sects~\ref{model_overview_text},~\ref{tides_winds}, and \ref{HJs}).

\appendix
\section{Computation of the libration and circulation periods}
\label{app1}
To compute the period of libration, we note that it corresponds to four times the time taken  to go from $\alpha=0$ to the maximum excursion $\alpha_{0}= \arcsin(E/\omega_{\rm p})$. In other words, the libration period is:
\begin{equation}
P_{\rm libr} = 4 \int_{0}^{\alpha_{0}} \left( \frac{dt}{d\alpha} \right) d\alpha = 4 \int_{0}^{\alpha_{0}} \frac{d\alpha}{\dot{\alpha}}, 
\end{equation}
where $\dot{\alpha}$ comes from equation~(\ref{first_integ}). In this way, we obtain
\begin{equation}
P_{\rm libr} =  \frac{4}{\omega_{\rm p} }\int_{0}^{\alpha_{0}} \frac{d\alpha}{\sqrt{\sin^{2} \alpha_{0}-\sin^{2} \alpha}}. 
\label{p_libr}
\end{equation}
Introducing a new angular variable $\xi$ as $\sin \xi \equiv \sin \alpha / \sin \alpha_{0}$, we have
\begin{equation}
d\alpha = \sin \alpha_{0} \frac{\cos \xi}{\sqrt{1- \sin^{2} \alpha_{0} \sin^{2} \xi}} d\xi. 
\end{equation}
Changing the variable of integration from $\alpha$ to $\xi$ in equation~(\ref{p_libr}), we have
\begin{equation}
P_{\rm libr} = \frac{4}{\omega_{\rm p}} \int_{0}^{\pi/2} \frac{d\xi}{\sqrt{1-\sin^{2}\alpha_{0} \sin^{2} \xi}} = \frac{4}{\omega_{\rm p}} K(\sin \alpha_{0}), 
\end{equation}
where $K(\gamma)$ is the complete elliptical integral of the first kind with $\gamma < 1$. For small values of $\gamma$, 
\begin{equation}
K(\gamma) \simeq \frac{\pi}{2} + \frac{\pi}{8} \frac{\gamma^{2}}{1-\gamma^{2}} - \frac{\pi}{16} \frac{\gamma^{4}}{1-\gamma^{2}} + \, \mbox{...}.
\end{equation}
In the case of the circulating solution $ E > \omega_{\rm p} $, $\dot{\alpha}$ has always the same sign, and we can again consider that the period is four times the time taken to go from $\alpha=0$ to $\alpha = \pi/2$:
\begin{equation}
P_{\rm circ} = \frac{4}{E} \int_{0}^{\pi/2} \frac{d\alpha}{\sqrt{1-(\omega_{\rm p}/E)^{2} \sin^{2} \alpha}} = \frac{4}{E} K \left(\frac{\omega_{\rm p}}{E} \right).  
\end{equation}

\section*{Acknowledgements}
The author  is grateful to an anonymous referee for a careful reading of the manuscript and several valuable comments that helped to improve this work. AFL acknowledges support by INAF/Frontiera through the "Progetti Premiali" funding scheme of the Italian Ministry of Education, University, and Research. The use of the web interface to the MESA code suite developed by C.~E.~Fields and F.~X.~Timmes is gratefully acknowledged. 
%
%

\begin{thebibliography}{99}

\bibitem[\protect\citeauthoryear{Amard, Palacios, Charbonnel, Gallet \& Bouvier}{2016}]{Amardetal16} 
Amard L., Palacios A., Charbonnel C., Gallet F., Bouvier J., 2016, A\&A, 587, A105

\bibitem[\protect\citeauthoryear{Applegate}{1989}]{Applegate89} 
Applegate J.~H., 1989, ApJ, 337, 865

\bibitem[\protect\citeauthoryear{Applegate}{1992}]{Applegate92} 
Applegate J.~H., 1992, ApJ, 385, 621

\bibitem[\protect\citeauthoryear{Applegate \& Patterson}{1987}]{ApplegatePatterson87} 
Applegate J.~H., Patterson J., 1987, ApJ, 322, L99

\bibitem[\protect\citeauthoryear{Barnes, Collier Cameron, Donati, James, Marsden \& Petit}{2005}]{Barnesetal05} 
Barnes J.~R., Collier Cameron A., Donati J.-F., James D.~J., Marsden S.~C., Petit P., 2005, MNRAS, 357, L1

\bibitem[\protect\citeauthoryear{Bonomo, et al.}{2017}]{Bonomoetal17} 
Bonomo A.~S., et al., 2017, A\&A, 602, A107

\bibitem[\protect\citeauthoryear{Borsa, et al.}{2015}]{Borsaetal15} 
Borsa F., et al., 2015, A\&A, 578, A64

\bibitem[\protect\citeauthoryear{Bours, et al.}{2016}]{Boursetal16} 
Bours M.~C.~P., et al., 2016, MNRAS, 460, 3873

\bibitem[Bradshaw, \& Hartigan(2014)]{BradshowHartigan14} 
{Bradshaw, S.~J., \& Hartigan, P.\ 2014, \apj, 795, 79 }

\bibitem[\protect\citeauthoryear{Brinkworth, Marsh, Dhillon \& Knigge}{2006}]{Brinkworthetal06} 
Brinkworth C.~S., Marsh T.~R., Dhillon V.~S., Knigge C., 2006, MNRAS, 365, 287

\bibitem[Brown et al.(2008)]{Brownetal08} 
{Brown, B.~P., Browning, M.~K., Brun, A.~S., et al.\ 2008, \apj, 689, 1354 } 

\bibitem[\protect\citeauthoryear{Browning}{2008}]{Browning08} 
Browning M.~K., 2008, ApJ, 676, 1262

\bibitem[\protect\citeauthoryear{Browning, Weber, Chabrier \& Massey}{2016}]{Browningetal16} 
Browning M.~K., Weber M.~A., Chabrier G., Massey A.~P., 2016, ApJ, 818, 189

\bibitem[\protect\citeauthoryear{Brun \& Browning}{2017}]{BrunBrowning17} 
Brun A.~S., Browning M.~K., 2017, LRSP, 14, 4

\bibitem[Brun et al.(2017)]{Brunetal17} 
{Brun, A.~S., Strugarek, A., Varela, J., et al.\ 2017, \apj, 836, 192}

\bibitem[\protect\citeauthoryear{Caligari, Moreno-Insertis \& Schussler}{1995}]{Caligarietal95} 
Caligari P., Moreno-Insertis F., Schussler M., 1995, ApJ, 441, 886

\bibitem[\protect\citeauthoryear{Collier Cameron, Donati \& Semel}{2002}]{CollierCameronetal02} 
Collier Cameron A., Donati J.-F., Semel M., 2002, MNRAS, 330, 699

\bibitem[\protect\citeauthoryear{Collier Cameron \& Donati}{2002}]{CollierCameronDonati02} 
Collier Cameron A., Donati J.-F., 2002, MNRAS, 329, L23

\bibitem[\protect\citeauthoryear{Collier Cameron \& Jardine}{2018}]{CollierCameronJardine18} 
Collier Cameron A., Jardine M., 2018, MNRAS, 476, 2542

\bibitem[\protect\citeauthoryear{Damiani \& Lanza}{2015}]{DamianiLanza15}  
Damiani C., Lanza A.~F., 2015, A\&A, 574, A39

\bibitem[\protect\citeauthoryear{Donati}{1999}]{Donati99} 
Donati J.-F., 1999, MNRAS, 302, 457

\bibitem[\protect\citeauthoryear{Donati \& Collier Cameron}{1997}]{DonatiCollierCameron97} 
Donati J.-F., Collier Cameron A., 1997, MNRAS, 291, 1

\bibitem[\protect\citeauthoryear{Donati, Collier Cameron \& Petit}{2003}]{Donatietal03} 
Donati J.-F., Collier Cameron A., Petit P., 2003, MNRAS, 345, 1187

\bibitem[\protect\citeauthoryear{Frasca \& Lanza}{2005}]{FrascaLanza05} 
Frasca A., Lanza A.~F., 2005, A\&A, 429, 309

\bibitem[Giles et al.(2018)]{Gilesetal17} 
{Giles, H.~A.~C., Collier Cameron, A., \& Haywood, R.~D.\ 2017, \mnras, 472, 1618}

\bibitem[\protect\citeauthoryear{Goldreich}{1966}]{Goldreich66} 
Goldreich P., 1966, AJ, 71, 1

\bibitem[\protect\citeauthoryear{Goldreich \& Peale}{1966}]{GoldreichPeale66} 
Goldreich P., Peale S., 1966, AJ, 71, 425

\bibitem[\protect\citeauthoryear{Hall}{1989}]{Hall89} 
Hall D.~S., 1989, SSRv, 50, 219

\bibitem[\protect\citeauthoryear{Hall}{1990}]{Hall90} 
Hall D.~S., 1990, NATO Advanced Science Institutes (ASI) Series C,  vol. 95, p. 319


\bibitem[\protect\citeauthoryear{Lanza}{2005}]{Lanza05} 
Lanza A.~F., 2005, MNRAS, 364, 238

\bibitem[\protect\citeauthoryear{Lanza}{2006}]{Lanza06} 
Lanza A.~F., 2006, MNRAS, 369, 1773

\bibitem[\protect\citeauthoryear{Lanza, Rodono \& Rosner}{1998}]{Lanzaetal98} 
Lanza A.~F., Rodono M., Rosner R., 1998, MNRAS, 296, 893

\bibitem[\protect\citeauthoryear{Lanza \& Rodon{\`o}}{1999}]{LanzaRodono99}
Lanza A.~F., Rodon{\`o} M., 1999, A\&A, 349, 887

\bibitem[\protect\citeauthoryear{Lanza \& Rodon{\`o}}{2001}]{LanzaRodono01} 
Lanza A.~F., Rodon{\`o} M., 2001, A\&A, 376, 165

\bibitem[\protect\citeauthoryear{Lanza \& Rodon{\`o}}{2004}]{LanzaRodono04}
Lanza A.~F., Rodon{\`o} M., 2004, AN, 325, 393

\bibitem[\protect\citeauthoryear{Lanza, Piluso, Rodon{\`o}, Messina \& Cutispoto}{2006}]{Lanzaetal06} 
Lanza A.~F., Piluso N., Rodon{\`o} M., Messina S., Cutispoto G., 2006, A\&A, 455, 595


\bibitem[\protect\citeauthoryear{Lazaridis, et al.}{2011}]{Lazaridisetal11} 
Lazaridis K., et al., 2011, MNRAS, 414, 3134

\bibitem[\protect\citeauthoryear{Lehtinen, Jetsu, Hackman, Kajatkari \& Henry}{2016}]{Lehtinenetal16} 
Lehtinen J., Jetsu L., Hackman T., Kajatkari P., Henry G.~W., 2016, A\&A, 588, A38

\bibitem[\protect\citeauthoryear{Marchioni, Guinan, Engle, Dowling Jones, Michail, Werner \& Ribas}{2018}]{Marchionietal18} 
Marchioni L., Guinan E.~F., Engle S.~G., Dowling Jones L., Michail J.~M., Werner G., Ribas I., 2018, RNAAS, 2, 179

\bibitem[\protect\citeauthoryear{Mardling \& Lin}{2002}]{MardlingLin02} 
Mardling R.~A., Lin D.~N.~C., 2002, ApJ, 573, 829

\bibitem[\protect\citeauthoryear{Marsh \& Pringle}{1990}]{MarshPringle90} 
Marsh T.~R., Pringle J.~E., 1990, ApJ, 365, 677

\bibitem[\protect\citeauthoryear{Matese \& Whitmire}{1983}]{MateseWhitmire83} 
Matese J.~J., Whitmire D.~P., 1983, A\&A, 117, L7

\bibitem[\protect\citeauthoryear{Meibom \& Mathieu}{2005}]{MeibomMathieu05} 
Meibom S., Mathieu R.~D., 2005, ApJ, 620, 970

\bibitem[\protect\citeauthoryear{Moreno-Insertis, Caligari \& Schuessler}{1995}]{Moreno-Insertisetal95} 
Moreno-Insertis F., Caligari P., Schuessler M., 1995, ApJ, 452, 894

\bibitem[\protect\citeauthoryear{Muneer, Jayakumar, Rosario, Raveendran \& Mekkaden}{2010}]{Muneeretal10} 
Muneer S., Jayakumar K., Rosario M.~J., Raveendran A.~V., Mekkaden M.~V., 2010, A\&A, 521, A36

\bibitem[\protect\citeauthoryear{Murray \& Dermott}{1999}]{MurrayDermott99} 
Murray C.~D., Dermott S.~F., 1999, Solar System Dynamics, Cambridge Univ. Press, Cambridge; Ch.~5 

\bibitem[\protect\citeauthoryear{Navarrete, Schleicher, Zamponi Fuentealba \& V{\"o}lschow}{2018}]{Navarreteetal18} 
Navarrete F.~H., Schleicher D.~R.~G., Zamponi Fuentealba J., V{\"o}lschow M., 2018, A\&A, 615, A81

\bibitem[\protect\citeauthoryear{Neff, O'Neal \& Saar}{1995}]{Neffetal95} 
Neff J.~E., O'Neal D., Saar S.~H., 1995, ApJ, 452, 879

\bibitem[Nelson et al.(2013)]{Nelsonetal13} 
{Nelson, N.~J., Brown, B.~P., Brun, A.~S., et al.\ 2013, \apj, 762, 73 }

\bibitem[Nelson et al.(2014)]{Nelsonetal14} 
{Nelson, N.~J., Brown, B.~P., Brun, A.~S., et al.\ 2014, \solphys, 289, 441 }

\bibitem[\protect\citeauthoryear{Ogilvie}{2014}]{Ogilvie14} 
Ogilvie G.~I., 2014, ARA\&A, 52, 171

\bibitem[\protect\citeauthoryear{Ogilvie \& Lin}{2007}]{OgilvieLin07} 
Ogilvie G.~I., Lin D.~N.~C., 2007, ApJ, 661, 1180

\bibitem[\protect\citeauthoryear{Ol{\'a}h, et al.}{2009}]{Olahetal09} 
Ol{\'a}h K., et al., 2009, A\&A, 501, 703

\bibitem[\protect\citeauthoryear{Paxton, et al.}{2011}]{Paxtonetal11} 
Paxton B., Bildsten L., Dotter A., Herwig F., Lesaffre P., Timmes F., 2011, ApJS, 192, 3

\bibitem[\protect\citeauthoryear{Perdelwitz, et al.}{2018}]{Peldelwitzetal18} 
Perdelwitz V., et al., 2018, A\&A, 616, A161

\bibitem[\protect\citeauthoryear{Phinney}{1992}]{Phinney92} 
Phinney E.~S., 1992, RSPTA, 341, 39

\bibitem[\protect\citeauthoryear{Pletsch \& Clark}{2015}]{PletschClark15} 
Pletsch H.~J., Clark C.~J., 2015, ApJ, 807, 18

\bibitem[\protect\citeauthoryear{Rodono, Lanza \& Catalano}{1995}]{Rodonoetal95} 
Rodono M., Lanza A.~F., Catalano S., 1995, A\&A, 301, 75

\bibitem[\protect\citeauthoryear{R{\"u}diger, Elstner, Lanza \& Granzer}{2002}]{Ruedigeretal02} 
R{\"u}diger G., Elstner D., Lanza A.~F., Granzer T., 2002, A\&A, 392, 605

\bibitem[\protect\citeauthoryear{R\"udiger \& Hollerbach}{2004}]{RuedigerHollerbach04} 
R\"udiger G., Hollerbach R., 2004, The Magnetic Universe: Geophysical and Astrophysical Dynamo Theory, Wiley-VCH Verlag, Weinheim 

\bibitem[\protect\citeauthoryear{Soderhjelm}{1980}]{Soderhjelm80} 
Soderhjelm S., 1980, A\&A, 89, 100

\bibitem[\protect\citeauthoryear{Vaccaro, Wilson, Van Hamme \& Terrell}{2015}]{Vaccaroetal15} 
Vaccaro T.~R., Wilson R.~E., Van Hamme W., Terrell D., 2015, ApJ, 810, 157

\bibitem[Viviani et al.(2018)]{Vivianietal18} 
{Viviani, M., Warnecke, J., K{\"a}pyl{\"a}, M.~J., et al.\ 2018, \aap, 616, A160 }

\bibitem[\protect\citeauthoryear{V{\"o}lschow, Schleicher, Perdelwitz \& Banerjee}{2016}]{Volschowetal16} 
V{\"o}lschow M., Schleicher D.~R.~G., Perdelwitz V., Banerjee R., 2016, A\&A, 587, A34

\bibitem[\protect\citeauthoryear{V{\"o}lschow, Schleicher, Banerjee \& Schmitt}{2018}]{Volschowetal18} 
V{\"o}lschow M., Schleicher D.~R.~G., Banerjee R., Schmitt J.~H.~M.~M., 2018, A\&A, 620, A42

\bibitem[Warnecke et al.(2018)]{Warneckeetal18} 
{Warnecke, J., Rheinhardt, M., Tuomisto, S., et al.\ 2018, \aap, 609, A51}

\bibitem[\protect\citeauthoryear{Watson \& Marsh}{2010}]{WatsonMarsh10} 
Watson C.~A., Marsh T.~R., 2010, MNRAS, 405, 2037

\bibitem[\protect\citeauthoryear{Wolff, Ray, Wood \& Hertz}{2009}]{Wolffetal09} 
Wolff M.~T., Ray P.~S., Wood K.~S., Hertz P.~L., 2009, ApJS, 183, 156

\end{thebibliography}




\bsp	
\label{lastpage}
\end{document}